\documentclass[twocolumn,a4paper,reprint,amssymb,aps,prd,groupedaddress,superscriptaddress,preprintnumbers,amsmath,nobibnotes,showpacs,nofootinbib]{revtex4-2}

\usepackage{graphicx}
\usepackage{epsf}
\usepackage{bm}

\usepackage{amsmath}
\usepackage[T1]{fontenc}
\usepackage{amsfonts}
\usepackage{amssymb}
\usepackage{epstopdf}
\usepackage{natbib}
\usepackage{color}
\usepackage[dvipsnames]{xcolor}
\usepackage{verbatim}
\usepackage{multirow}
\usepackage{physics}
\usepackage{pifont}
\usepackage{comment}
\usepackage{bm}
\usepackage{microtype}
\usepackage[colorlinks=true, linkcolor=blue , citecolor=blue, urlcolor=blue]{hyperref}
\usepackage{xcolor}
\usepackage{amsmath}
\usepackage[capitalize]{cleveref}
\usepackage[normalem]{ulem}
\usepackage{enumitem}
\usepackage{lipsum}
\usepackage{epstopdf}
\usepackage{etoolbox}

\DeclareUnicodeCharacter{2212}{-}
\newcommand{\bea}{\begin{eqnarray*}}
\newcommand{\eea}{\end{eqnarray*}}
\newcommand{\be}{\begin{equation}}
\newcommand{\ee}{\end{equation}}
\newcommand{\ba}{\begin{eqnarray}}
\newcommand{\ea}{\end{eqnarray}}

\begin{document}

\preprint{YITP-24-119}

\title{Unexplored regions in teleparallel $f(T)$ gravity: Sign-changing dark energy density}

\author{\"{O}zg\"{u}r Akarsu}
\email{akarsuo@itu.edu.tr}
\affiliation{Department of Physics, Istanbul Technical University, Maslak 34469 Istanbul, T\" urkiye}

\author{Bilal Bulduk}
\email{bulduk21@itu.edu.tr}
\affiliation{Department of Physics, Istanbul Technical University, Maslak 34469 Istanbul, T\" urkiye}
\affiliation{Department\;of\;Physics,\;Gebze\;Technical\;University,\;Gebze,\;41400\;Kocaeli,\;T\" urkiye}

\author{Antonio De Felice}
\email{antonio.defelice@yukawa.kyoto-u.ac.jp}
\affiliation{
Center for Gravitational Physics and Quantum Information, Yukawa Institute for Theoretical Physics, Kyoto University, 606-8502, Kyoto, Japan
}

\author{Nihan Kat{\i}rc{\i}}
\email{nkatirci@dogus.edu.tr}
\affiliation{Department of Electrical and Electronics Engineering Do\u gu\c s University \"Umraniye, 34775 Istanbul, T\" urkiye}

\author{N. Merve Uzun}
\email{uzunmer@itu.edu.tr}
\affiliation{Department of Physics, Istanbul Technical University, Maslak 34469 Istanbul, T\" urkiye}

\begin{abstract}
While teleparallel $f(T)$ gravity has shown considerable potential in addressing cosmological tensions, such as the $H_0$ and $S_8$ discrepancies, we explore previously overlooked solution spaces within this framework that hold further promise. Specifically, we examine the case where the customary assumption of a strictly positive effective dark energy (DE) density---natural in general relativity---may not apply, offering new possibilities. Focusing on the exponential infrared model \( f(T) = T e^{T_*/T} \), where \( T_* \) is a characteristic torsion scale, we investigate cosmological solutions parametrized by the dimensionless parameter \( \beta = T_*/T_0 \) with $T_0$ the present-day torsion scalar. This parameter uniquely determines the present-day matter density parameter \( \Omega_{\rm m0} \), and its sign plays a crucial role in characterizing deviations from the standard $\Lambda$ cold dark matter ($\Lambda$CDM) expansion history. We elaborate on the structural asymmetry between the positive- and negative-$\beta$ branches: while the positive branch ($\beta_{+}$) leads to dynamics with modest departures from $\Lambda$CDM, the negative branch ($\beta_{-}$) yields more pronounced and nontrivial deviations at cosmological scales. We discuss that, despite these deviations, the negative-$\beta$ branch can remain consistent with local gravity constraints through an effective chameleon-like mechanism---wherein high-density environments naturally suppress deviations from the teleparallel equivalent of general relativity. We extend our analysis by examining the model in the context of dynamical DE. Ensuring consistency with cosmic microwave background (CMB) data, we find that the widely studied $\beta_{+}$ case exhibits phantom behavior, while the previously overlooked $\beta_{-}$ case---sufficient to avoid instabilities or ghosts---features a sign-changing DE density that transitions smoothly from negative to positive values at redshift $z_{\dagger} \sim 1.5$, consistent with recent approaches to alleviating multiple cosmological tensions. Though the sign-changing DE density in the $f(T)$ model leads to a larger-than-expected enhancement, we further extend the analysis by incorporating a cosmological constant, $\Lambda$. This extension, $f(T) \rightarrow f(T) + 2\Lambda$, broadens the solution space consistent with the SH0ES $H_0$ measurement while maintaining consistency with CMB power spectra. Additionally, it introduces richer phenomenological possibilities, including the potential moderation or cessation of cosmic acceleration at very low redshifts, aligning with recent observational analyses, such as those from DESI BAO data. Our findings also suggest that existing $f(T)$ models, as well as background-equivalent $f(Q)$ models, should be revisited in light of the novel theoretical insights presented here.

\end{abstract}
\maketitle

\section{Introduction}
In the teleparallel equivalent of general relativity (TEGR), constructed from a torsion-based approach to gravity, the gravitational part of the Lagrangian density is described by the torsion scalar $T$, replacing the torsionless Levi-Civita connection of the standard curvature-based formulation with the vanishing-curvature Weitzenb\"ock connection~\cite{Aldrovandi:2013wha}. The generalization of the gravitational Lagrangian density $T$ to an arbitrary function $f(T)$~\cite{Ferraro:2006jd,Ferraro:2008ey}, in a similar spirit to $f(R)$ theory~\cite{Copeland:2006wr,Sotiriou:2008rp,DeFelice:2010aj,Nojiri:2010wj,Clifton:2011jh,Capozziello:2011et,Bamba:2012cp,Nojiri:2017ncd}, has been extensively investigated from various perspectives, generating considerable interest over the past two decades. A key factor driving this interest is that field equations of $f(T)$ gravity are second order, unlike those of $f(R)$ gravity, which are fourth order, although Ostrogradsky instabilities can be avoided. The $f(T)$ theory, whose initial formulation was not locally Lorentz invariant, includes extra degrees of freedom that are absent in the standard general relativity (GR). The covariant formulation of the $f(T)$ gravity, which is safe from local Lorentz symmetry violation~\cite{Li:2010cg,Sotiriou:2010mv}, has been explored in Refs.~\cite{Arcos:2004tzt, Krssak:2015oua,Golovnev:2017dox}.  However, it should be noted that ongoing debates remain, particularly concerning the number of degrees of freedom in the theory~\cite{Ong:2013qja,Maluf:2018coz,Bejarano:2019fii,Golovnev:2020zpv,Golovnev:2021lki,Golovnev:2021omn}. A recent perturbative analysis~\cite{Danieli:2025mov} has confirmed that in $f(T)$ gravity there are only two propagating degrees of freedom in the gravity sector, regardless of whether one assumes a maximally symmetric Friedmann-Lema\^{i}tre-Robertson-Walker (FLRW) background~\cite{Dent:2010nbw,Chen:2010va,Izumi:2012qj,Li:2011wu,Golovnev:2018wbh,Sahlu:2019bug,Bahamonde:2022ohm,Hohmann:2020vcv} or the less symmetric Bianchi type~I spacetime. Moreover, under the spatially flat FLRW assumption, the $f(T)$ background equations coincide exactly with those of $f(Q)$ gravity (where $Q$ is nonmetricity scalar)~\cite{Jarv:2018bgs}, so all background‐level analyses performed in the $f(Q)$ context apply directly to $f(T)$ models with the same functional form. We refer readers to Refs.~\cite{Cai:2015emx,Golovnev:2018wbh,Krssak:2018ywd,Bahamonde:2021gfp} for a comprehensive review on teleparallel $f(T)$ gravity.

As with other modified gravity theories, $f(T)$ gravity models, effective at late times in the Universe, have been investigated to determine whether they can provide an interpretation of the present-day accelerated cosmic expansion driven by torsional effects~\cite{Bengochea:2008gz, Linder:2010py}; see also Ref.~\cite{Hohmann:2017jao} for a dynamical system analysis of a generic $f(T)$ model. Detailed analyses of the dynamics of these models, based on various choices of the $f(T)$ function, have revealed that the effective dark energy (DE) component arising from the extra torsional terms can exhibit quintessence and phantom regimes, as well as phantom divide line (PDL) crossings~\cite{Wu:2010av,Karami:2010bys,Bamba:2010wb,Cardone:2012xq}. In recent years, it has been reported that simple phantom models mitigate the $H_0$ (Hubble constant) tension, while models featuring PDL crossing---first suggested through a model now referred to as DMS20~\cite{DiValentino:2020naf}---offer even greater relief for the $H_0$ tension. A more recent analysis of the DMS20 model~\cite{DiValentino:2020naf}, as presented in Refs.~\cite{Adil:2023exv,Specogna:2025guo}, interpreted as an embodiment of omnipotent DE, has not only confirmed its effectiveness in mitigating the $H_0$ tension but also revealed that the model’s ability to attain negative density values for $z \gtrsim 2$ and resemble to a negative cosmological constant at higher redshifts, along with the PDL crossing at $z \sim 0.1$, plays a crucial role in alleviating the tension. This aligns with the findings from the $\Lambda_{\rm s}$ cold dark matter ($\Lambda_{\rm s}$CDM) model~\cite{Akarsu:2021fol, Akarsu:2022typ, Akarsu:2023mfb,Escamilla:2025imi}---which suggests a rapid (smooth or abrupt) transition of the Universe from anti-de Sitter (AdS) vacua to de Sitter (dS) vacua in the late Universe at redshift $z_\dagger \sim 2$, as conjectured through the graduated dark energy (gDE) model~\cite{Akarsu:2019hmw}---shown to be promising in simultaneously addressing major cosmological tensions such as the $H_0$ and $S_8$ (growth parameter) tensions, as well as several other less significant tensions. We refer readers, without claiming to be exhaustive, to Refs.~\cite{Sahni:2002dx,Vazquez:2012ag,BOSS:2014hwf,Sahni:2014ooa,BOSS:2014hhw,DiValentino:2017rcr,Mortsell:2018mfj,Poulin:2018zxs,Capozziello:2018jya,Wang:2018fng,Banihashemi:2018oxo,Dutta:2018vmq,Banihashemi:2018has,Akarsu:2019ygx,Li:2019yem,Visinelli:2019qqu,Ye:2020btb,Perez:2020cwa,Akarsu:2020yqa,Calderon:2020hoc,Ye:2020oix,DeFelice:2020cpt,Paliathanasis:2020sfe,Bonilla:2020wbn,Acquaviva:2021jov,Bag:2021cqm,Bernardo:2021cxi,Escamilla:2021uoj,DiGennaro:2022ykp,Akarsu:2022lhx,Bernardo:2022pyz,Ong:2022wrs,Tiwari:2023jle,Malekjani:2023ple,Alexandre:2023nmh,Gomez-Valent:2023uof,Medel-Esquivel:2023nov,Anchordoqui:2023woo,Anchordoqui:2024gfa,Gomez-Valent:2024tdb,Bousis:2024rnb,Manoharan:2024thb,Wang:2024hwd,Colgain:2024ksa,Tyagi:2024cqp,Yadav:2024duq,Toda:2024ncp,Dwivedi:2024okk,Anchordoqui:2024dqc,Gomez-Valent:2024ejh,Souza:2024qwd,Mukherjee:2025myk,Giare:2025pzu,Keeley:2025stf,Soriano:2025gxd,Wang:2025dtk,Bouhmadi-Lopez:2025ggl,Bouhmadi-Lopez:2025spo,Paraskevas:2024ytz,Akarsu:2025ijk,Efstratiou:2025xou,Gonzalez-Fuentes:2025lei,Hogas:2025ahb,Mishra:2025goj} for further studies that explore DE models with negative energy density values, often consistent with a negative (AdS-like) cosmological constant, particularly for $z \geq 1.5-2$, and aimed at addressing major cosmological tensions.  For recent reviews on cosmological tensions and discrepancies, see Refs.~\cite{DiValentino:2021izs,Perivolaropoulos:2021jda,Abdalla:2022yfr,DiValentino:2022fjm,Vagnozzi:2023nrq,Akarsu:2024qiq,CosmoVerse:2025txj}. Recently, the Dark Energy Spectroscopic Instrument (DESI) baryon acoustic oscillations (BAO) data, both with and without the inclusion of Planck cosmic microwave background (CMB)~\cite{Planck:2018vyg} and Type Ia Supernovae (SNIa) data~\cite{Brout:2022vxf,Rubin:2023ovl,DES:2024tys}, have provided more than $2\sigma$ evidence for dynamical DE when using the Chevallier-Polarski-Linder (CPL) parametrization~\cite{DESI:2024mwx}. Additionally, nonparametric reconstructions of the DE density based on DESI BAO data, combined with other datasets, suggest the possibility of vanishing or negative DE densities for $z \gtrsim 1.5-2$~\cite{DESI:2024aqx,Escamilla:2024ahl}, a phenomenon also observed in pre-DESI BAO data, particularly the Sloan Digital Sky Survey (SDSS) BAO data~\cite{Escamilla:2023shf,Sabogal:2024qxs,Escamilla:2024ahl}.

The abrupt $\Lambda_{\rm s}$CDM model~\cite{Akarsu:2021fol, Akarsu:2022typ, Akarsu:2023mfb} mentioned above, which offers one of the most economical frameworks for addressing major cosmological tensions while accommodating a wide range of datasets with only one additional parameter---the redshift of the proposed mirror AdS-to-dS transition---was initially conjectured phenomenologically based on observational findings from the gDE model~\cite{Akarsu:2019hmw}. However, realizing this hypothesized rapid transition of the cosmological constant in the late Universe seems theoretically challenging, particularly given that it involves a shift from negative to positive values. Nevertheless, the model's remarkable phenomenological success has sparked theoretical interest, leading to recent advances that propose plausible mechanisms for this transition. For instance, Refs.~\cite{Anchordoqui:2023woo,Anchordoqui:2024gfa,Anchordoqui:2024dqc,Soriano:2025gxd} propose that Casimir forces from fields within the dark dimension scenario could drive the abrupt late-time mirror AdS-to-dS transition. While the AdS swampland conjecture suggests that such a transition might be improbable, these studies indicate that an AdS-to-dS shift in the late Universe is theoretically achievable under specific conditions, leading to a stringy realization of the $\Lambda_{\rm s}$CDM model. Similarly, Refs.~\cite{Akarsu:2024qsi,Akarsu:2024eoo} show that rapid (smooth or abrupt) late-time mirror AdS-to-dS transitions, or analogous DE dynamics, can be effectively realized through a particular Lagrangian in a type II minimally modified gravity theory called V cold dark matter (VCDM)~\cite{DeFelice:2020eju,DeFelice:2020cpt}. This realization, referred to as $\Lambda_{\rm s}$VCDM, involves an auxiliary scalar field with a two-segmented linear potential and differing in perturbation dynamics from the GR-based $\Lambda_{\rm s}$CDM model, achieves an even better fit to observational data~\cite{DeFelice:2020cpt}. The recently proposed phantom  
$\Lambda_{\rm s}$ cold dark matter (Ph-$\Lambda_{\rm s}$CDM) model~\cite{Akarsu:2025gwi,Akarsu:2025dmj} explores smooth cosmological transitions mediated by a minimally coupled scalar field with a hyperbolic tangent potential. By modeling dark energy as a phantom field within this framework, the model naturally induces a late-time AdS-to-dS transition—first in the scalar potential, followed by a corresponding transition in the energy density—without encountering stability issues. Among other alternatives for realizing a late-time AdS-to-dS transition, or DE dynamics with similar characteristics, are approaches involving an overall change of the metric signature in GR~\cite{Alexandre:2023nmh}, bimetric gravity~\cite{Dwivedi:2024okk}, and Horndeski gravity~\cite{Tiwari:2023jle}, although this is by no means a complete list. These recent advances underscore that a late-time AdS-to-dS transition---or more broadly, a transition from negative to positive DE density---may be more feasible than initially anticipated. Thus, it may be time to reconsider the conventional assumption that DE density must remain positive---a perspective rooted in GR and in the historical view that negative DE densities have not been considered cosmologically relevant and have frequently been associated with instabilities. These assumptions, however, are familiar within the GR framework and may not extend to effective DE models emerging from modified gravity frameworks. Furthermore, embedding the $\Lambda_{\rm s}$CDM model within a theoretical framework by defining a specific Lagrangian elevates it from a phenomenological proposal to a fully predictive model, universally applicable and even beyond cosmological scales, as demonstrated in $\Lambda_{\rm s}$VCDM~\cite{Akarsu:2024qsi,Akarsu:2024eoo}. Developing DE models with such dynamics from theories with well-defined Lagrangians is therefore essential and strongly motivates the work presented in this paper. Thus, revisiting established modified gravity theories without imposing a strict positivity constraint on DE density may open promising avenues for addressing cosmological tensions. As a case study, we present a previously studied $f(T)$ model that aligns with the perspective we outlined above and yields solution spaces---exhibiting an effective DE component with the desired characteristics, particularly in the late Universe---that may have been overlooked due to the conventional assumptions. Although the nature of the physical degrees of freedom for $f(T)$ is still an open issue, the model we discuss does fulfill the stability conditions for the perturbations during the background evolution in which we are interested.

In this paper, focusing on a particular model known as \textit{exponential infrared teleparallel gravity}~\cite{Awad:2017yod,Hashim:2020sez,Hashim:2021pkq}, described by the function $f(T)=Te^{T_*/T}$---where \( T_* \) is the characteristic torsion scale from which one can define the dimensionless parameter \( \beta = T_*/T_0 \), with \( T_0 \) representing the present-day value of the torsion scalar relevant in a cosmological context---and examining all possible solution spaces, we reveal unexplored solution regions that showcase how dynamics featuring a transition from negative to positive energy density values in the late Universe---specifically,  a sign-changing behavior in the effective DE density--- can be achieved within torsion-based $f(T)$ gravity models. In Sec.~\ref{sec:teleparallel}, we briefly present the necessary background for $f(T)$ theory. In Sec.~\ref{sec:6p}, we discuss the widely studied and the uncharted regions that predict effective torsional DE featuring phantom behavior as well as the uncharted regions in which the effective torsional DE exhibits nontrivial sign-changing behavior in its energy density. As a distinguishing feature, this model is based on six parameters like the standard $\Lambda$CDM such that $\beta$ is not a free parameter but is determined by the present-day energy density of matter ($\Omega_{\rm m0}$) in the context of FLRW cosmology. After presenting the extensive theoretical discussion, in Sec.~\ref{sec:cosm}, we explore the viable cosmologies, particularly concentrating on the late-time accelerating cosmic expansion. The studies so far adhered to $\beta>0$  case, viz., the positive exponent ($\beta_+$), excluding negative one, and obtained an effective DE whose density parameter is below the PDL. However, we show that $\beta<0$ case, viz., negative exponent ($\beta_-$) conversely generates an effective DE whose energy density attains negative values beyond a certain redshift. On top of this interesting feature, $\beta_-$ is also a sufficient condition to avoid instabilities/ghosts, independently of the value dynamics of $T$ on any background, whereas $\beta_+$ is not. Then, we further point out that $f(T)$ gravity might be a potential candidate for realizing an AdS-dS transition, not rapidly but in a comparably smooth way, through a DE component effectively arising from the extra torsional terms due to $f-T$ modification. Although not realizing a mirror AdS-dS transition, our findings showcase the $f(T)$ gravity's capability on the way to integrating $\Lambda_{\rm s}$CDM model into a theoretical framework. Then, in Sec.~\ref{sec:lambda}, we extend the viable cosmologies by introducing a cosmological constant as $f(T)\rightarrow f(T)+2\Lambda$ with $f(T)=Te^{\beta T_0/T}$ whose $\beta=0$ limit is equivalent to the standard $\Lambda$CDM model on a spatially flat FLRW background. We show that even this simplest modification to the original form of the function widens the scope of the possibilities in the cosmological context to a considerable extent. In Sec.~\ref{sec:final}, we draw our conclusions.

\section{Cosmology in Teleparallel Description of Gravity}
\label{sec:teleparallel}
A general spacetime is a four-dimensional differentiable manifold $\mathcal{M}$, where the tangent space at each point is a Minkowski spacetime. We introduce four linearly independent vector fields, $e^{a}{}_{\mu}$, known as vierbeins or tetrads, defined on this smooth manifold. Here, Greek indices $(\mu, \nu, ...)$ correspond to the general spacetime coordinates, while Latin indices $(a, b, ...)$ denote the tangent space coordinates, both running from $0$ to $3$~\cite{Aldrovandi:2013wha,Cai:2015emx,Krssak:2018ywd,Bahamonde:2021gfp,Golovnev:2018wbh}.
To ensure a nondegenerate metric, the vierbeins must satisfy the orthonormality conditions $e^{\mu}{}_a e^{a}{}_{\nu}=\delta^{\mu}{}_{\nu}$ and $e^b{}_{\mu}e^{\mu}{}_a=\delta^b{}_a$.
Consequently, the Lorentzian metric tensor of the spacetime can be expressed as
\begin{equation}
g_{\mu\nu}=\eta_{ab}e^{a}{}_{\mu}e^{b}{}_{\nu},
\label{eq:metricT}
\end{equation}
where $\eta_{ab}=\text{diag}(-1,1,1,1)$ is the metric tensor of Minkowski spacetime in Cartesian coordinates.

In teleparallel gravity, the gravitational interaction is attributed to spacetime torsion rather than curvature, requiring a connection distinct from the Levi-Civita connection used in the standard GR. In particular, we employ the general affine connection expressed in terms of the tetrad field and the spin connection $\omega^a{}_{\mu b}$ that accounts for inertial effects~\cite{Golovnev:2017dox}, as follows:
\begin{align}
\Gamma^\sigma{}_{\mu\nu} = e^{\sigma}{}_{a}\left(\partial_\mu e^a{}_\nu + \omega^a{}_{\mu b} e^b{}_\nu\right),
\end{align}
which is generically nonsymmetric and gives rise to the torsion tensor
\begin{align}
T^{\sigma}{}_{\mu\nu} &\equiv \Gamma^{\sigma}{}_{\mu\nu} - \Gamma^{\sigma}{}_{\nu\mu} \nonumber \\
&= e^{\sigma}{}_{a} \left( \partial_\mu e^a{}_\nu - \partial_\nu e^a{}_\mu + \omega^a{}_{\mu b} e^b{}_\nu - \omega^a{}_{\nu b} e^b{}_\mu \right).
\label{eq:torsion}
\end{align}
An analogy can be drawn between the torsion tensor $T^{\sigma}{}_{\mu\nu}$ in teleparallel gravity and the Riemann curvature tensor, $R^{\sigma}{}_{\mu\nu\alpha}$, in the conventional representation of GR; see Ref.~\cite{Golovnev:2023qll} for further details.
 Then, we define the spacetime-indexed superpotential tensor as
\begin{equation}
    S_{\sigma\mu\nu}=\frac{1}{2}\left(T_{\sigma\mu\nu}+T_{\nu\mu\sigma}+T_{\mu\sigma\nu}\right)+g_{\sigma\mu}T_{\nu}-g_{\sigma\nu}T_{\mu},
\label{eq:superpotential}
\end{equation}
whose first three terms form contortion tensor, $K_{\mu\sigma\nu}$, which is antisymmetric in the lateral indices and $T_{\mu} \equiv T^{\sigma}{}_{\mu\sigma}$ is the torsion vector. Contracting it with the torsion tensor yields the torsion scalar, viz., $T=\frac{1}{2}S_{\sigma\mu\nu}T^{\sigma\mu\nu}$, which can also be expressed as
\begin{align}
T=\frac{1}{4}T_{\sigma\mu\nu}T^{\sigma\mu\nu}+\frac{1}{2}T_{\sigma\mu\nu}T^{\mu\sigma\nu}-T_{\mu}T^{\mu}.
\label{eq:torsionscalar}
\end{align}
Consequently, the action of the $f(T)$ gravity reads~\cite{Golovnev:2017dox}
\begin{equation}
   \mathcal{S}=\int {\rm d}^4 x\; \vert \vert e \vert \vert \left[-\frac{1}{2\kappa} f(T)+\mathcal{L}_{\rm{m}}\right],
\label{eq:action}
\end{equation}
where $\vert \vert e \vert \vert={\rm det} (e^a{}_{\mu})=\sqrt{-g}$ with $g$ being the determinant of $g_{\mu\nu}$, $\kappa=8\pi G_{\rm N}$ with $G_{\rm N}$ being the standard Newtonian constant, and $\mathcal{L}_{{\rm m}}$ is the Lagrangian density of material fields. Here and throughout the paper, we work in units such that the speed of light equals unity, $c=1$. We note that the particular case $f(T)=T$ reduces to TEGR~\cite{Aldrovandi:2013wha}.

This study investigates the cosmological applications of $f(T)$ gravity. To this end, we adopt the following diagonal vierbein configuration:
\begin{equation}
e^{a}{}_{\mu}={\rm diag} [1,a(t),a(t),a(t)],
\label{eq:vierbeinF}
\end{equation}
which corresponds to the FLRW spacetime metric with flat spatial sections, given by
\begin{equation}
\begin{aligned}
\label{eqn:RWmetric}
{\rm d}s^2=&-{\rm d}t^2+a(t)^2\left({\rm d}x^2+{\rm d}y^2+{\rm d}z^2\right),
\end{aligned}
\end{equation}
 where $a(t)$ is the scale factor and $t$ denotes cosmic time. Substituting Eq.~\eqref{eq:vierbeinF} into Eq.~\eqref{eq:torsionscalar} with Eq.~\eqref{eq:torsion}, we obtain the torsion scalar as
\begin{equation}
\label{eq:THreln}
T=6H^2,
\end{equation}
where $H=\frac{\dot{a}}{a}$ is the Hubble parameter. In the covariant formulation of $f(T)$ gravity, the spin connection $\omega^a{}_{\mu b}$ must be treated as an independent variable and varied in the action, yielding the antisymmetric part of the tetrad field equations~\cite{Krssak:2015oua,Hohmann:2019nat}. This ensures local Lorentz invariance of the theory. However, for the diagonal FLRW spacetime metric tetrad adopted in~\cref{eq:vierbeinF}, constructed in Cartesian coordinates, the spin connection components can be safely set to zero since it is a pure gauge connection. As a result, the antisymmetric part of the field equations is trivially satisfied in our case, and the covariant and noncovariant formulations coincide at the background level. We suppose that all types of matter distributions (viz., the usual cosmological sources such as radiation, baryons, etc.) are perfect fluids with no peculiar velocities. Accordingly, by expanding at first order the action~\eqref{eq:action} on the FLRW background, we find the modified Friedmann equations for an arbitrary function of the torsion scalar, viz.,  $f(T)$, as follows~\cite{Danieli:2025mov}:
\begin{align}  
\label{eq:rho}
3H^2&=\tfrac{1}{2}\left(T+f-2Tf_T\right)+\kappa\rho, \\
-2\Dot{H}-3H^2&=2\Dot{H}(f_T+2 T f_{TT}-1) \nonumber \\ &\quad-\tfrac{1}{2}(T+f-2Tf_T)+\kappa p,
\label{eq:p}
\end{align}
where $f_T={\rm d} f/{\rm d} T$ and $f_{TT}={\rm d}^2f/{\rm d} T^2$. Here $\rho$ and $p$ denote the energy density and pressure of the perfect fluid, respectively.\footnote{In the presence of multiple components, $\rho$ and $p$ denote the total energy density and total pressure, respectively, of the standard matter fields (namely radiation, baryons, dark matter, and vacuum energy).} This expression of the dynamical equations allows us to immediately write a general expression for the effective contribution from $T$ to the energy density and pressure as
\begin{align}
    \kappa\rho_{\rm T}&\equiv\tfrac{1}{2}\left(T+f-2Tf_T\right)\,,\label{eq:rhoT_gen}\\
    \kappa p_{\rm T} &\equiv 2\Dot{H}(f_T+2 T f_{TT}-1) -\tfrac{1}{2}(T+f-2Tf_T)\,.\label{eq:pT_gen}
\end{align}
And, it is immediate to verify that Eqs.~\eqref{eq:rhoT_gen} and~\eqref{eq:pT_gen} satisfy by construction the continuity equation $\dot{\rho}_{\rm T}+3H(\rho_{\rm T}+p_{\rm T})=0$. We note that the effective torsional contributions $\rho_{\rm T}$ and $p_{\rm T}$, given in Eqs.~\eqref{eq:rhoT_gen} and~\eqref{eq:pT_gen}, vanish identically in two cases: the trivial choice $f(T)=T$ (TEGR), and the nontrivial square-root extension $f(T)=T+\alpha \sqrt{T}$, where $\alpha$ is an arbitrary constant. In both models, the background Friedmann equations reduce exactly to those of standard GR. However, in the square-root extension, the full field equations (and hence the perturbation equations) acquire additional $\sqrt{T}$-dependent terms from the variation of the action~\eqref{eq:action} (see Ref.~\cite{Hohmann:2017jao}). A detailed analysis allowing small deviations around these two GR limits has been conducted within the framework of $f(Q)$ gravity, whose Friedmann equations coincide with those of $f(T)$ gravity for a specifically chosen connection, in order to investigate the dynamical stability of the Universe across its evolutionary eras~\cite{Guzman:2024cwa}. The study shows that the models accommodate stable radiation and matter eras, as well as stability during the late-time, DE–dominated era.\footnote{While both GR limits yield the same background evolution in this regime, the stability properties differ depending on how the cosmological constant is realized. In the geometric DE case---where an effective cosmological constant arises from the torsion/nonmetricity modification itself as a constant offset in the GR limit, e.g., $\Lambda_{\rm eff} \equiv \kappa \rho_{\rm T} = \text{const.}$, in line with Eq.~(50) of Ref.~\cite{Guzman:2024cwa}, corresponding to the introduction of a bare cosmological constant in GR---the model remains stable. In contrast, when the cosmological constant is introduced explicitly as vacuum energy in the matter sector, the model enters a marginally stable phase during the late-time, DE-dominated era. See Ref.~\cite{Guzman:2024cwa} for further details.} Therefore, it is reasonable to expect that any viable cosmological model within the $f(T)$ gravity framework should satisfy $f \sim T$ or $f \sim T + \alpha \sqrt{T}$ for $T \gg T_0$, where $T_0$ denotes the present-day value of the torsion scalar---in order to preserve the successful description of the early Universe, including the recombination and pre-recombination eras, as well as standard big bang nucleosynthesis. Likewise, to recover the standard $\Lambda$CDM behavior near the present epoch, the model should satisfy $f \sim T + \text{const.}$ or $f \sim T + \alpha \sqrt{T} + \text{const.}$ for $T \sim T_0$. Teleparallel gravity theories constructed from $f(T)$ functions that asymptotically match these GR-compatible forms in the respective limits can robustly reproduce the well-established cosmic epochs. In this context, investigating the intermediate redshift regime---where multiple matter components are subdominant and deviations intrinsic to the theory may emerge---becomes particularly relevant for probing the theory’s distinctive phenomenology. Having established this theoretical background and motivation, we now proceed to examine a specific model of $f(T)$ gravity within the context of a spatially homogeneous and isotropic universe.

\section{Exponential Infrared Teleparallel Cosmology}
\label{sec:6p}
\begin{figure}[t!]
\par
\begin{center}
\vspace*{0.5mm}\includegraphics[trim =0mm  0mm 0mm 0mm, clip, width=0.48\textwidth]{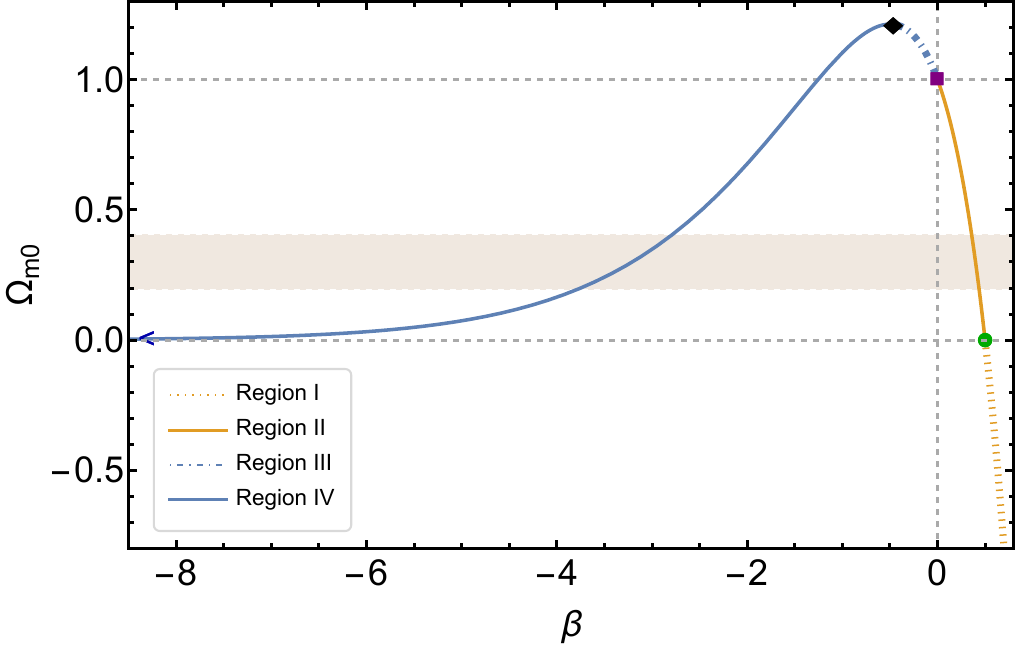}
\end{center}
\caption{$\Omega_{\rm m0}$ vs $\beta$ plotted by using Eq.~\eqref{eq:constraint}. The curve is separated into four regions as described in Sec.~\ref{fig:SolutionRegions}: the dotted orange part (Region I) is the new region that covers negative values of the present-day density parameter of matter, i.e.,\ $\Omega_{\rm m0}<0$. The solid orange part (Region II) is the widely studied region~\cite{Awad:2017yod,Hashim:2020sez,Hashim:2021pkq}, which corresponds to $0<\Omega_{\rm m0}< 1$, because reasonable $\Omega_{\rm m0}$ values from the observational point of view shown by the wheat-colored band lie in this region. The dot-dashed blue curve (Region III) is another new region leading to $\Omega_{\rm m0}$ values larger than unity. The solid blue curve (Region IV) is an overlooked region in the literature even though the observationally reasonable $\Omega_{\rm m0}$ values can be obtained in this region as well. Some special points are  $(\beta=-1/2,\,\Omega_{\rm m0}=2/\sqrt{e}=1.21306)$ represented by black diamond, $(\beta \to -\infty, \,\Omega_{\rm m0} \to 0)$ by blue triangle, $(\beta=1/2, \,\Omega_{\rm m0}=0)$ by green circle and $(\beta=0, \,\Omega_{\rm m0}=1)$ by purple square.}
\label{fig:evo_beta_2}
\end{figure}

In this work, we begin with a specific model known as exponential infrared teleparallel gravity, originally proposed in Ref.~\cite{Awad:2017yod}, and described by the functional form
\begin{equation} 
\label{eq:model0}
    f(T)=Te^{T_*/T},
\end{equation}
where \( T_* \) represents a characteristic torsion scale. In the context of cosmological model building, it is convenient to express this function in terms of the present-day value of the torsion scalar, denoted as $T_0$. Defining the dimensionless parameter $\beta=T_*/T_0$, we can equivalently rewrite Eq.~\eqref{eq:model0} in a form better suited for cosmological applications:
\begin{equation} 
\label{eq:model}
    f(T)=Te^{\beta\, T_0/T}.
\end{equation}
Considering that pressureless matter follows the local energy-momentum conservation law, as in GR, viz., $\dot{\rho}_{\rm m}+3H\rho_{\rm m}=0$~\cite{Cai:2015emx}, and substituting Eq.~\eqref{eq:model} into Eq.~\eqref{eq:rho} along with the relation~\eqref{eq:THreln}, we obtain the following independent Friedmann equation and the constraint equation, respectively:
\begin{align} 
\label{eq:fried1}
    \left(\frac{H^2}{H_0^2}-2\beta \right)e^{\beta H_0^2/H^2}&= \Omega_{\rm m0}(\beta)\,(1+z)^3,\\
    \Omega_{\rm m0}(\beta)&=(1-2 \beta) e^{\beta}\,,
    \label{eq:constraint}
\end{align}
where $\Omega_{\rm m0}=\kappa\rho_{\rm m0}/(3H_0^2)$  is the present-day matter density parameter. The density parameter is defined as $\Omega=\rho/\rho_{\rm cr}$, with the critical energy density of the Universe given by $\rho_{\rm{cr}}=3 H^2/\kappa$. The redshift  is related to the scale factor as $z=a_0/a-1$, where $a_0$ denotes the present-day value of the scale factor. Here and throughout this paper, a subscript $0$ attached to any quantity denotes its present-day value (i.e., at $z = 0$). Note that Eq.~\eqref{eq:constraint} is derived by applying the condition $H(z=0)=H_0$ to Eq.~\eqref{eq:fried1}. This constraint equation indicates that the model does not introduce any new free parameters compared to the standard, namely, six-parameter based, $\Lambda$CDM model. See Fig.~\ref{fig:evo_beta_2}, which explicitly shows that $\beta$  distinguishes different regions near a cosmic phase transition point (the critical point $\beta=0$) taking the place of the present-day matter density, $\Omega_{\rm m0}$. We also note that the requirement of a positive present-day matter density, $\Omega_{\rm m0} > 0$, imposes an upper bound on the parameter, namely $\beta < \tfrac{1}{2}$. For now, we assume $\beta<1/2$ in line with the aim of the following discussion, but we will leave this assumption while extensively exploring the theoretical capabilities of the model.

To gain deeper insight into the relation between $\beta$ and $\Omega_{\rm m0}$, the constraint equation~\eqref{eq:constraint} can be inverted analytically as
\begin{align} \label{lambert}
\beta = \frac{1}{2} + W_k\left( -\frac{\Omega_{\rm m0}}{2 \sqrt{e}} \right),
\end{align}
where $W_k$ denotes the Lambert $W$ function, with two real-valued branches $k=0$ and $k=-1$ yielding distinct solutions. This inversion reveals that for any given value of $\Omega_{\rm m0}$ within the physically plausible range $0<\Omega_{\rm m0} \leq 1$, there are typically two admissible solutions for $\beta$: one positive and one negative---excluding the special case $\Omega_{\rm m0}=1$, for which one of the solutions is $\beta=0$, rather than positive. These two branches are expected to yield distinct cosmological behaviors, despite reproducing the same present-day matter density parameter. Let us consider an illustrative example that makes this immediately evident: the special case $\beta = 0$ corresponds to the teleparallel equivalent of general relativity, where $f(T) = T$, yielding the Einstein--de Sitter universe---that is, a matter-dominated FLRW universe with $\Omega_{\rm m0} = 1$ and a vanishing cosmological constant. However, the converse does not hold: alongside $\beta = 0$, the condition $\Omega_{\rm m0} = 1$ also admits a second solution at $\beta \approx -1.256$, corresponding to a nontrivial, torsionally modified cosmology. This case highlights how allowing negative values of $\beta$ opens up qualitatively new branches of evolution, even when the present-day matter content is identical. It is interesting to note that there is no \textit{a priori} lower bound on $\beta$, and the entire negative $\beta$ domain corresponds to solutions within the physically allowed range $\Omega_{\rm m0} > 0$.  More specifically, for the interval $0 < \Omega_{\rm m0} < 1$, where two real solutions for $\beta$ exist, the positive branch spans the narrow interval $(0, 1/2)$, whereas the negative branch extends over the much broader range $(-\infty, -1.256)$, unbounded from below, as can be observed in Fig.~\ref{fig:evo_beta_2}. For instance, a typical observational value such as $\Omega_{\rm m0} = 0.3$ yields two distinct solutions for $\beta$: a mildly positive value, $\beta = 0.399$, and a significantly negative one $\beta = -3.207$. This illustrates that the asymmetry between the two branches is not merely a matter of sign. For the same $\Omega_{\rm m0}$, the magnitude of the negative solution is nearly an order of magnitude larger than that of the positive one. Consequently, the $\beta < 0$ solution branch entails more substantial departures from the standard $\Lambda$CDM expansion history, leading to enhanced sensitivity to initial conditions and, therefore, potentially tighter observational constraints on the model. Interestingly, small negative values of $\beta$ can still be realized if one allows $\Omega_{\rm m0} > 1$. In this extended regime, both branches of the Lambert $W$ function yield negative $\beta$ values on either side of $\beta = -1/2$, where $\Omega_{\rm m0}$ reaches its maximum value of $1.213$, corresponding to the branch cut, see Fig.~\ref{fig:evo_beta_2}. Specifically, the principal branch ($k = 0$) gives solutions in the range $-1/2 < \beta < 0$, while the secondary branch ($k = -1$) yields values in the range $-1.256 < \beta < -1/2$.

Throughout this work, we aim to illuminate the role of the negative $\beta$ solutions, which have been largely overlooked in comparison to the positive branch that is widely examined in the literature. The presence of these structurally distinct branches---degenerate in $\Omega_{\rm m0}$ but dynamically inequivalent---becomes apparent upon closer examination of the model, supplementing our earlier discussion on the viability of the $f(T)$ model under consideration. Equation~\eqref{eq:model} shows that our model approaches the standard TEGR limit ($f(T) = T$) in the high-torsion regime $T \gg T_0$, regardless of the sign of $\beta$. On the other hand, the behavior near the present epoch of the Universe---where $T \sim T_0$---warrants a more systematic investigation of both positive and negative $\beta$ scenarios. Since $\beta \ll 1$ in the positive branch, the model yields $f(T) \sim T + \beta T_0$; that is, it approaches the TEGR limit with a cosmological constant-like term---thereby reproducing a standard $\Lambda$CDM-like cosmology. However, the form of the $f(T)$ function prevents us from straightforwardly drawing conclusions about the negative branch: while $|\beta| \ll 1$ does not hold, $\mathcal{O}(|\beta|) \sim 1$ does, as discussed in the previous paragraph. Consequently, the Taylor expansion of the $f(T)$ function at linear order breaks down.

To further motivate this investigation, we note that the structural asymmetry between the $\beta > 0$ and $\beta < 0$ branches is expected to manifest in the modified Hubble function. To make this explicit, we recast Eqs.~\eqref{eq:fried1} and~\eqref{eq:constraint} into the form
\begin{equation}
   \label{eq:hubble-recast}
E(z)^2 = (1 - 2\beta)\, e^{\beta} (1 + z)^3\, e^{-\beta / E(z)^2} + 2\beta,
\end{equation}
or, equivalently,
\begin{equation}
E(z)^2 = \Omega_{\rm m0}(\beta)\,(1+z)^3\,e^{-\beta / E(z)^2} + 1 - \Omega_{\rm m0}(\beta)\,e^{-\beta},
\end{equation}
where \( E(z) = H(z)/H_0 \) is the normalized Hubble parameter. For comparison, the standard \(\Lambda\)CDM background evolution is given by
\begin{equation}
E(z)^2 = \Omega_{\rm m0} (1+z)^3 + 1 - \Omega_{\rm m0},
\end{equation}
neglecting radiation. Both models involve the same number of background-level parameters: \(\{H_0, \Omega_{\rm m0}\}\) for \(\Lambda\)CDM and \(\{H_0, \beta\}\) for the model under consideration. However, for any given value of \(\Omega_{\rm m0}>0\), there exist two corresponding values of \(\beta\), denoted by \(\beta_{k=0}\) and \(\beta_{k=-1}\), arising from the two real branches of the Lambert \(W\) function in Eq.~\eqref{lambert}---except the branch cut corresponding to \(\Omega_{\rm m0}=1.21306\) for which the \(\beta\) values in both branches coincide as \(\beta_{k=0}=\beta_{k=-1}=-1/2\). As a result, while the expansion history, $E(z)$, in \(\Lambda\)CDM is uniquely determined by \(\Omega_{\rm m0}\), the same value of \(\Omega_{\rm m0}\) in the present model corresponds to two dynamically distinct branches.

We observe that the deviations from the standard \(\Lambda\)CDM model manifest in two key aspects of the normalized Hubble expansion function: \textbf{(i)} There exists a constant term, \(2\beta\), contributing to \(E(z)^2\), analogous to the cosmological constant in \(\Lambda\)CDM. However, unlike in \(\Lambda\)CDM, where this contribution is fixed as \(\Omega_{\Lambda0} = 1 - \Omega_{\rm m0}\), the corresponding term in the present model takes the form \(\Omega_{\Lambda0}(\beta) = 1 - \Omega_{\rm m0}(\beta)\, e^{-\beta}\), allowing for a broader range of behavior. Depending on the sign and magnitude of \(\beta\), \(\Omega_{\Lambda0}(\beta)\) can take either positive or negative values for a given value of $\Omega_{\rm m0}$. \textbf{(ii)} There exists a dynamical term contributing to \(E(z)^2\), which, however, differs from \(\Omega_{\rm m0}(1+z)^3\)---the standard redshift evolution of the matter content in \(\Lambda\)CDM---by acquiring a \(\beta\)-dependent dynamical modulation through the exponential factor \(e^{-\beta / E(z)^2}\). As a result, it takes the form \(\Omega_{\rm m0}(\beta)(1+z)^3\,e^{-\beta / E(z)^2}\), introducing a structurally distinct contribution to the expansion history. Crucially, both of these departures depend explicitly on \(\beta\), and therefore differ between the two branches even when \(\Omega_{\rm m0}\) is held fixed. In other words, for the same present-day matter density \(\Omega_{\rm m0}\), the model yields two distinct contributions to the effective cosmological constant–like term, as well as two distinct redshift evolutions of the dynamical term, corresponding to the \(\beta_{k=0}\) and \(\beta_{k=-1}\) branches.

\begin{figure}[t!]
\par
\begin{center}
\vspace*{0.5mm}\includegraphics[trim =0mm  0mm 0mm 0mm, clip, width=0.48\textwidth]{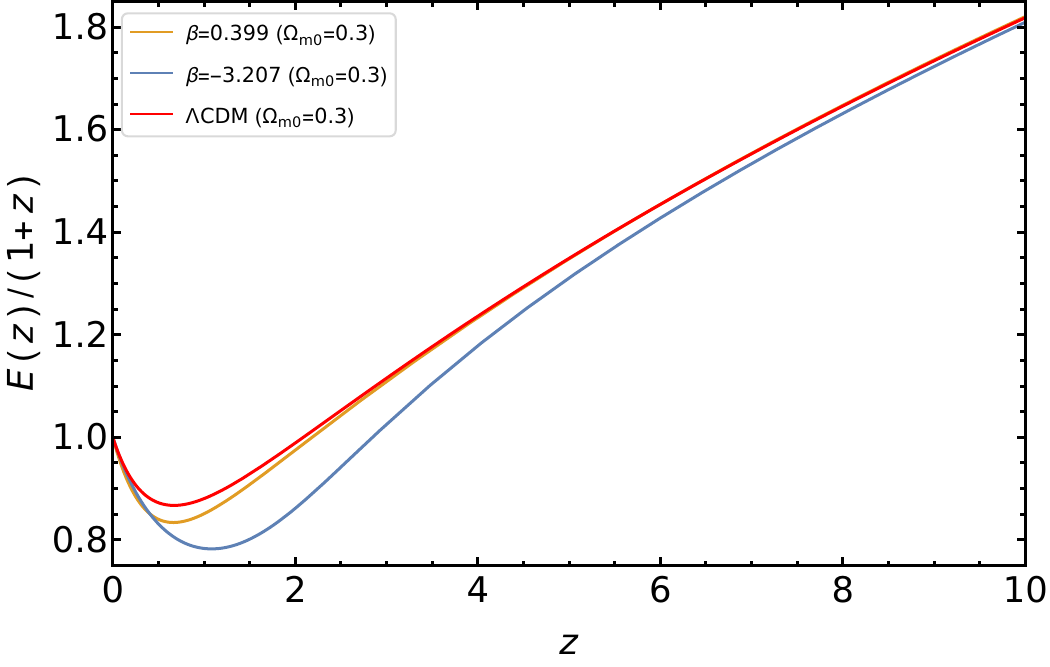}
\end{center}
\caption{Evolution of the normalized Hubble parameter scaled by $(1+z)$, i.e., $E(z)/(1+z)$, as a function of redshift $z$, for the standard $\Lambda$CDM model (red) and for the present teleparallel gravity model with $\Omega_{\rm m0}=0.3$. The two branches of the model correspond to $\beta = 0.399$ (orange, positive branch) and $\beta = -3.207$ (blue, negative branch). All three curves converge to the Einstein--de Sitter behavior at high redshift in the matter-dominated era. While the positive branch shows only mild deviations from $\Lambda$CDM, the negative branch exhibits substantial departures in the post-matter-dominated regime.}
\label{fig:Ez}
\end{figure}

The \(\beta < 1/2\) assumption we have made for satisfying the condition \(\Omega_{\rm m0} > 0\) renders the matter contribution in the considered model physically viable. Furthermore, requiring a physically viable expansion history---specifically \(\dot{H} < 0\), i.e., a monotonically increasing \(H(z)\) with redshift---ensures that both branches recover the standard early-Universe behavior. In the high-redshift limit \(H \gg H_0\), corresponding to \(z \gg 0\), both branches asymptotically approach the Einstein--de Sitter background, as follows:
\begin{equation}
E(z)^2 \simeq \Omega_{\rm m0}(\beta)\, (1+z)^3.
\end{equation}
However, \(\beta\)-dependent departures from $\Lambda$CDM emerge in the post-matter-dominated era. As discussed above, these deviations are moderate for $\beta>0$ branch, while becoming substantial in the $\beta<0$ branch. This asymmetry becomes evident when evaluating the model for a common value of \(\Omega_{\rm m0} = 0.3\), which corresponds to \(\beta = 0.399\) in the positive branch and \(\beta = -3.207\) in the negative branch. In these cases, the modified Hubble functions become
\begin{align}
E(z)^2 &= 0.3\,(1+z)^3\,e^{-0.399 / E(z)^2} + 0.798 \;\, (\beta > 0), \\
E(z)^2 &= 0.3\,(1+z)^3\,e^{+3.207 / E(z)^2} - 6.414 \;\, (\beta < 0),
\end{align}
to be compared with the standard \(\Lambda\)CDM expression
\begin{equation}
E(z)^2 = 0.3\,(1+z)^3 + 0.7.
\end{equation}
In the positive branch, relative to \(\Lambda\)CDM, the present-day density parameter associated with the cosmological constant-like term is slightly larger: \(\Omega_{\Lambda0} = 0.798\) compared to \(\Omega_{\Lambda0} = 0.7\) in \(\Lambda\)CDM. Moreover, the exponential modulation term features a small negative exponent, evolving from approximately unity at early times to \(\sim 0.671\) today. This leads to a slightly faster decline in the dynamical term compared to the standard \((1+z)^3\) scaling in $\Lambda$CDM. In the negative branch, by contrast, \(\beta\) is not only negative but also an order of magnitude larger in absolute value than in the positive branch. This is directly reflected in the cosmological constant-like term, which now takes significantly negative value: \(\Omega_{\Lambda0} = -6.414\). Furthermore, the exponential modulation term becomes highly sensitive to variations in \(E(z)\), increasing from approximately unity at early times to \(\sim 24.70\) today. As a result, the dynamical term undergoes a substantial slowdown in its scaling relative to the standard \((1+z)^3\) behavior in $\Lambda$CDM, marking a significant departure from standard matter evolution. In \cref{fig:Ez}, we illustrate the evolution of the normalized Hubble parameter scaled by $(1+z)$, i.e., \(E(z)/(1+z)\), for both branches, assuming a common value of \(\Omega_{\rm m0} = 0.3\), alongside the \(\Lambda\)CDM prediction for comparison. As for the future behavior of the Universe---which is not shown in the figure---the positive branch exhibits a gradual fading of the dynamical term, and it is straightforward to verify that the Universe asymptotically approaches a de Sitter phase, $E(z)\rightarrow\sqrt{2\beta}$. In stark contrast, the negative branch evolves toward a state in which  \(E(z)\) eventually vanishes, signaling a future Minkowski phase---consistent with Eq.~\eqref{eq:fried1}, which admits an asymptotic solution with vanishing right-hand side only if \(H(z) \rightarrow 0\) for \(\beta < 0\) as $z\rightarrow-1$.

These results reveal a fundamental asymmetry between the positive- and negative-$\beta$ branches that is intrinsic to the model’s structure, highlighting the nontrivial, sign-sensitive role that the parameter $\beta$ plays in shaping the cosmic expansion history. Considering a plausible value of $\Omega_{\rm m0} = 0.3$, we have shown that both branches yield cosmologically interesting and relevant solutions. That said, while the $\beta > 0$ sector is generally more robust and effectively introduces only small corrections to the standard $\Lambda$CDM scenario, the $\beta < 0$ sector is phenomenologically richer but yields substantial deviations from $\Lambda$CDM in the post-matter-dominated epoch---leading to a highly nontrivial and less conventional expansion history that, although realizing accelerated expansion at the present epoch, asymptotically approaches a Minkowski spacetime in the future. Furthermore, as illustrated in~\cref{fig:evo_beta_2}, the model exhibits markedly greater sensitivity to variations in $\Omega_{\rm m0}$ within the negative-$\beta$ branch. This is because the condition $\Omega_{\rm m0} > 0$ restricts the positive-$\beta$ branch to a narrow range $0 < \beta < 1/2$, while the negative branch accommodates the entire semi-infinite interval $-\infty < \beta < 0$. As a result, small changes in $\Omega_{\rm m0}$ translate into disproportionately large shifts in the dynamics of the negative-$\beta$ models, highlighting the need for a more delicate treatment of this sector. Since the Hubble function~\eqref{eq:fried1} is defined only implicitly, preventing an analytic reconstruction of the expansion history in terms of redshift, we treat $H$ as a parametric variable and proceed with a systematic numerical exploration of both branches. A complete dynamical analysis of the model, particularly in the $\beta < 0$ sector, would provide further insights into its global evolution; however, this is reserved for future work. In the remainder of the present study, we instead focus on disentangling the evolution of the actual source---dust composed of baryons and cold dark matter---from the additional torsional terms, which we identify as an effective DE component. Although closed-form solutions are not available, we employ a combination of numerical methods and implicit function analysis, supported by illustrative figures, to examine the kinematics and dynamics of the effective DE across the full range of $\beta$.

To facilitate this investigation, we first recast Eq.~\eqref{eq:fried1} using the constraint in Eq.~\eqref{eq:constraint}, which allows us to express the redshift as an explicit function of the Hubble parameter, as follows:
\begin{equation}  
\label{redshift}
z(H)=e^{\frac\beta3 (H_0^2/H^2-1)}\!\left[\frac{H^2/H_0^2-2\beta}{1-2\beta}\right]^{1/3}-1,
\end{equation}
provided that $\beta \neq 1/2$. And, using the derivative of Eq.~\eqref{redshift} with respect to $H$, we derive another important relation given by
\begin{align}
   \frac{{\rm d} H}{{\rm d}\mathcal{N}}&=-(1+z)\frac{{\rm d}H}{{\rm d} z},\nonumber\\
    &=-\frac{3}{2}\,\frac{H\left(1-2 \beta  H_0^2/H^2\right)}{1 -\beta  H_0^2/H^2+2 \beta ^2 H_0^4/H^4},\label{eq:dH_dN}
     \end{align}
where $\mathcal{N}\equiv\ln(a/a_0)$ is the e-fold variable. Note that the denominator in Eq.~\eqref{eq:dH_dN} never vanishes for any real value of $\beta$ or $H_0/H$.\footnote{Setting the denominator in Eq.~\eqref{eq:dH_dN} to zero transforms it into a quadratic equation with the variable redefined as $x \equiv H_0^2/H^2$. For real values of $\beta$, the equation always has a negative discriminant, $\Delta = -7 \beta^2$, indicating two complex (nonreal) roots, which are not valid solutions for $H_0^2/H^2$.  Hence, Eq.~\eqref{eq:dH_dN} is free from singularities.} Next, the Ricci scalar, constructed using the metric $g_{\mu\nu}$, is given by
\begin{align}
   R &= 12H^2-6H\, (1+z) \,\frac{dH}{dz},\nonumber\\
   &=12H^2-\frac{9 H^2\left(1-2\beta H_0^2/H^2\right)}{1-\beta H_0^2/H^2+2\beta^2H_0^4/H^4}\,. \label{eq:Ricci_scalar}
\end{align}
We also calculate the deceleration parameter, $q\equiv -1-\frac{1}{H}\frac{{\rm d} H}{{\rm d} \mathcal{N}}$, as follows:
\begin{equation} \label{eq:q}
q=-1+\frac{3}{2} \,\frac{1-2\beta H_0^2/H^2}{1-\beta H_0^2/H^2+2\beta^2H_0^4/H^4}.
\end{equation}
For $\beta = 0$ (marked by the purple square in Fig.~\ref{fig:evo_beta_2}), we find $q = 1/2$, corresponding to the deceleration parameter of a matter-dominated universe in TEGR, which describes the Einstein--de Sitter universe, as there is no bare cosmological constant introduced. In the following section, Fig.~\ref{fig:evo_H_dec} illustrates the behavior of the Hubble and deceleration parameters with respect to redshift for different values of the parameter $\beta$, specifically $\Omega_{\rm m0}$. 
See the straight purple line in the bottom panel of Fig.~\ref{fig:evo_H_dec} plotted for $\beta = 0$; the Einstein--de Sitter universe ($q = 1/2$) is the critical point. 

\subsection{New solution regions along with known regions}
\label{fig:SolutionRegions}

It can be deduced from Eq.~\eqref{eq:fried1} that the model exhibits distinct characteristic behaviors depending on the values of the parameter $\beta$. These behaviors can be classified into the following regions and special points, as illustrated in Fig.~\ref{fig:evo_beta_2}:
\newcounter{qcounter}
\begin{list}
{\bfseries{}Region I~}
{
\usecounter{qcounter}
}
\item $\left(\beta>1/2 \right)$: the present-day matter density parameter is negative, i.e., $\Omega_{\rm m0}<0$ as required by Eq.~\eqref{eq:constraint}, and this region is represented by the dotted orange curve in Fig.~\ref{fig:evo_beta_2}. As $\beta$ increases, $\Omega_{\rm m0}$ diverges to negative infinity ($\Omega_{\rm m0} \to -\infty$). For this region, Eq.~\eqref{eq:fried1} implies that the Hubble parameter is constrained by $H^2<2\beta H_0^2$. Analysing the evolution of $H(z)$, we find that in the past, as $z \to \infty$, $H \to 0$ with $q\to-1$, corresponding to a Minkowski spacetime, and in the far future, as $z \to -1$, $H^2\to 2\beta H_0^2$ with $q\to-1$, representing a dS universe, as demonstrated in Fig.~\ref{fig:evo_H_dec}. 
\end{list}
\begin{list}
{\bfseries{}Point I~}
{
\usecounter{qcounter}
}
\item $\left(\beta=1/2\right)$:
this case corresponds to $\Omega_{\rm m0}=0$ and is represented by the green point in Fig.~\ref{fig:evo_beta_2}. This special point leads to an empty universe, i.e., a zero present-day density parameter of matter. Nevertheless, a dS background with $H^2=H_0^2$ ($q=-1$) still emerges due to the modification in the spacetime geometry from $f(T)$, even though no bare cosmological constant is introduced, as shown by the green line in Fig.~\ref{fig:evo_H_dec}.
\end{list}
\begin{list}
{\bfseries{}Region II~}
{
\usecounter{qcounter}
}
\item $\left(0<\beta<1/2\right)$:
the solid orange curve in Fig.~\ref{fig:evo_beta_2} represents this region, which includes the well-known case already discussed in the literature~\cite{Hashim:2020sez,Hashim:2021pkq}. In this region, the present-day matter density parameter lies within the interval $0<\Omega_{\rm m0}<1$, enabling a reasonable range of $\Omega_{\rm m0}$, as indicated by the wheat-colored band in Fig.~\ref{fig:evo_beta_2}. Here, Eq.~\eqref{eq:fried1} constrains the Hubble parameter to $H^2>2\beta H_0^2$. Analysing the evolution of $H(z)$, we find that in the past, as $z \to \infty$, $H \to \infty$ with $q\to 1/2$, corresponding to a matter-dominated universe, and in the far future, as $z \to -1$, $H^2\to 2\beta H_0^2$ with $q\to-1$, representing a dS universe, as illustrated in Fig.~\ref{fig:evo_H_dec}. This region yields a viable torsional DE model featuring phantom behavior, which has been thoroughly investigated in Sec.~\ref{sec:phantom}.

\begin{figure}[t!]
\par
\begin{center}
\hspace*{1.mm}\includegraphics[trim =0mm  0mm 0mm 0mm, clip, width=0.48\textwidth]{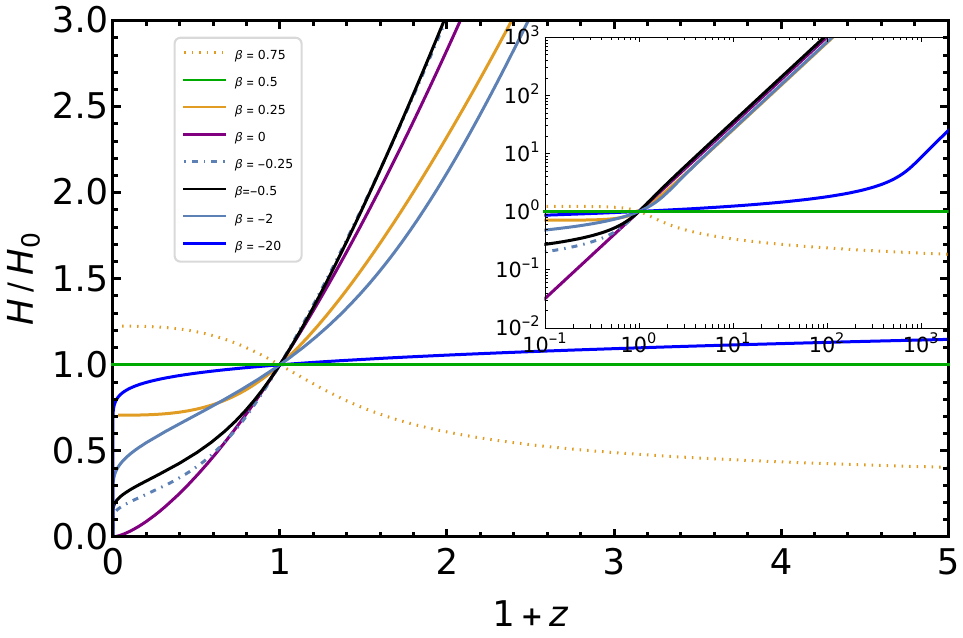}
\medskip
\hspace*{-0.7mm}\includegraphics[trim =0mm  0mm 0mm 0mm, clip, width=0.493\textwidth]{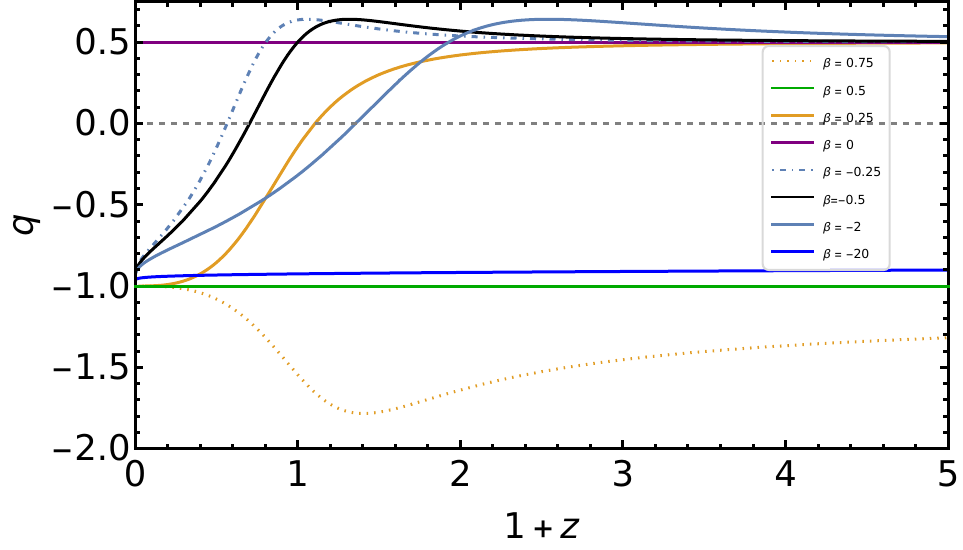}
\end{center}
\caption{\textbf{Top panel:} Hubble parameter $H(z)$ scaled by $H_0$. \textbf{Bottom panel:} deceleration parameter $q(z)$ with respect to $1+z$ for some $\beta$ values chosen according to the regions presented in Fig~\ref{fig:evo_beta_2} where colors are encoded with those used for regions.}
\label{fig:evo_H_dec}
\end{figure}

\end{list}

\begin{list}
{\bfseries{}Point II~}
{
\usecounter{qcounter}
}
\item $\left(\beta=0  \right)$: this case is denoted by a purple square marker corresponding to $\Omega_{\rm m0}=1$ in Fig.~\ref{fig:evo_beta_2}. Here, the extra terms arising from $f(T)-T$ vanish, and thus the $f(T)$ regime transitions a matter-dominated TEGR, i.e., an Einstein--de Sitter universe with $q=1/2$, as shown in Fig~\ref{fig:evo_H_dec}. This represents a critical point due to the absence of accelerated expansion.
\end{list}

\begin{list}
{\bfseries{}Region III~}
{
\usecounter{qcounter}
}
\item $\left(-1/2<\beta<0\right)$: 
in this case, the present-day density parameter of matter lies within the range $1<\Omega_{\rm m0}<2/\sqrt{e}$, which is represented by the blue dot-dashed curve and is being discussed for the first time in the literature. In this region, as demonstrated in the bottom panel of the Fig.~\ref{fig:evo_H_dec}, the transition from deceleration to acceleration happens in the future ($z_{\rm tr}<0$); therefore, this case cannot account for the present-day accelerated expansion of the Universe.
\end{list}

\begin{list}
{\bfseries{}Point III~}
{
\usecounter{qcounter}
}
\item $\left(\beta=-1/2\right)$: 
in this case, shown by the black diamond marker in Fig.~\ref{fig:evo_beta_2}, the present-day matter density parameter reaches its maximum value, $\Omega_{\rm m0}=2/\sqrt{e}$, within the model under consideration. The dynamics observed are similar to those of Region~III, as illustrated in Fig~\ref{fig:evo_H_dec}.
\end{list}

\begin{list}
{\bfseries{}Region IV~}
{
\usecounter{qcounter}
}
\item $\left(\beta< -1/2 \right)$: 
in this case, represented by the blue curve, the present-day matter density parameter lies within the range $0<\Omega_{\rm m0}<2/\sqrt{e}$. Notably, this region also includes the observationally reasonable $\Omega_{\rm m0}$ values indicated by the wheat-colored band, though it has been largely overlooked in cosmological analyses to date. This region yields an effective DE model with a sign-change in its density formed from torsional terms. This second viable model is thoroughly investigated in Sec.~\ref{sec:signchange}. As $\beta$ becomes more negative (corresponding to smaller $\Omega_{\rm m0}$), the transition redshift from deceleration to acceleration shifts further into the past.\footnote{As we will also see later on, the change of sign happens for $|\beta| H_0^2/H^2=-W_{-1}(-2^{-1}{\rm e}^{-1/2})-1/2$, $W_{k}$ being the Lambert-W function, leading to $H\approx 0.892|\beta|^{1/2} H_0$.}
\end{list}

Concerning the last three cases ($\beta < 0$), in the far future, i.e., for $z\to-1$, the solution of Eq.~\eqref{eq:fried1} leads to $H\to0$. In the same limit, we observe that ${\rm d}H/{\rm d}z\to\infty$, which might raise concerns about a potential singularity. However, from Eq.~\eqref{eq:dH_dN}, it becomes evident that $H\to0$ and ${\rm d}H/{\rm d}\mathcal{N}\to0$, indicating a finite Minkowski limit.\footnote{For example, from Eq.~\eqref{eq:Ricci_scalar}, we can see that the Ricci scalar built from the metric $g_{\mu\nu}$ also tends to vanish in this Minkowski limit. This is because this invariant, as well as others constructed from the metric tensor, contains terms involving $(1+z)\,{\rm d}H/{\rm d}z$ or equivalently ${\rm d}H/{\rm d}\mathcal{N}$, which remain finite (along with higher derivatives such as ${\rm d}^2H/{\rm d}\mathcal{N}^2$, etc.).}

\subsection{The effective gravitational constant, the signature of $f_T$, and extended regions}

We leave for future work the question of the stability of the solutions against linear inhomogeneous and anisotropic perturbation, yet we now discuss $f_T$ and its contribution to effective Newtonian constant by adapting it into a variable form considering the evolution of the matter perturbations in the sub-horizon limit which is governed by 
\begin{equation} 
\label{eq:matpert}
\ddot{\delta}_{\rm m} + 2H\dot{\delta}_{\rm m}-\frac{4\pi G_{\rm N}}{f_T}\rho_{\rm m}\delta_{\rm m}=0\,,
\end{equation}
where $\delta_{\rm m}=\delta \rho_{\rm m}/\rho_{\rm m}$ is the matter density contrast. Thus, the linear structure formation during the matter-dominated era is affected both by the modified background evolution through $H$ and by the modulation of the effective Newtonian constant for the cosmological perturbations through~\cite{Golovnev:2018wbh}
\begin{equation} \label{Gcosmo}
\mathcal{G}_{\rm eff} \equiv \frac{G_{\rm N}}{f_T}.
\end{equation}
The last term in Eq.~\eqref{eq:matpert} should be negative to ensure that matter does cluster around galaxies and structure is formed properly. In this regard, if we consider dark matter as yet an unknown component of $\Omega_{\rm m0}$ as in the standard scenario, the positivity of $f_T$ guarantees the attractive nature of gravity, having $\mathcal{G}_{\rm eff}>0$, then matter clusters around the galaxies.

Now we would like to discuss the signature of $f_T$ without a detailed investigation on scalar and tensor perturbations since it is an important quantity in $f(T)$ theory, particularly, at the perturbation level. The derivative of the $f(T)$ function given in Eq.~\eqref{eq:model} with respect to the torsion scalar $T$ reads
 \begin{align}  \label{fT-gen}
f_T&=e^{\beta T_0/T}\left(1-\beta T_0/T \right). 
\end{align}
It can be easily deduced from Eq.~\eqref{fT-gen} that $\beta<0$ is a sufficient condition to have $f_T>0$, independently of the value dynamics of $T$ on any background, including the FLRW spacetime, which may be seen from Eq.~\eqref{fT} as well. Conversely, when $\beta>0$, one needs to ensure that dynamically the Universe never entered through an era during which $f_T<0$. Using the relation~\eqref{eq:THreln} in the Eq.~\eqref{fT-gen} assuming FLRW spacetime metric, we obtain
 \begin{align}
f_T=e^{\beta H_0^2/H^2}\left(
1-\beta H_0^2/H^2 \right). \label{fT}
\end{align}
Hence, on FLRW spacetime assumption, we notice from Eq.~\eqref{fT} that $\beta <0$ case including Regions III and IV and Point III ($\beta=-1/2$) satisfies $f_T>0$ irrespective of the evolution of Hubble parameter.
For Point II $(\beta=0)$, we have $f_T=1$. In the $\beta>0$ case, Region II and Point I $(\beta=1/2)$ ensure the positivity of $f_T$; for the former, the Hubble parameter is already bounded as $H^2>2\beta H_0^2$ and for the latter, we have $f_T=\sqrt{e}/2$. Lastly, for Region I where $H^2$ is bounded as $0<H^2<2 \beta H_0^2$, there are two possibilities: the interval $0<H^2<\beta H_0^2$ leads to $f_T<0$, whereas the interval $\beta H_0^2<H^2<2\beta H_0^2$ implies $f_T>0$. In other words, in Region~I, $\Omega_{\rm m0}$ is always negative but $f_T$ takes both positive and negative values according to the evolution of the Hubble parameter.

\begin{figure}[t!]
\par
\begin{center}
\hspace*{1.5mm}\includegraphics[trim = 0mm  0mm 0mm 0mm, clip, width=0.48\textwidth]{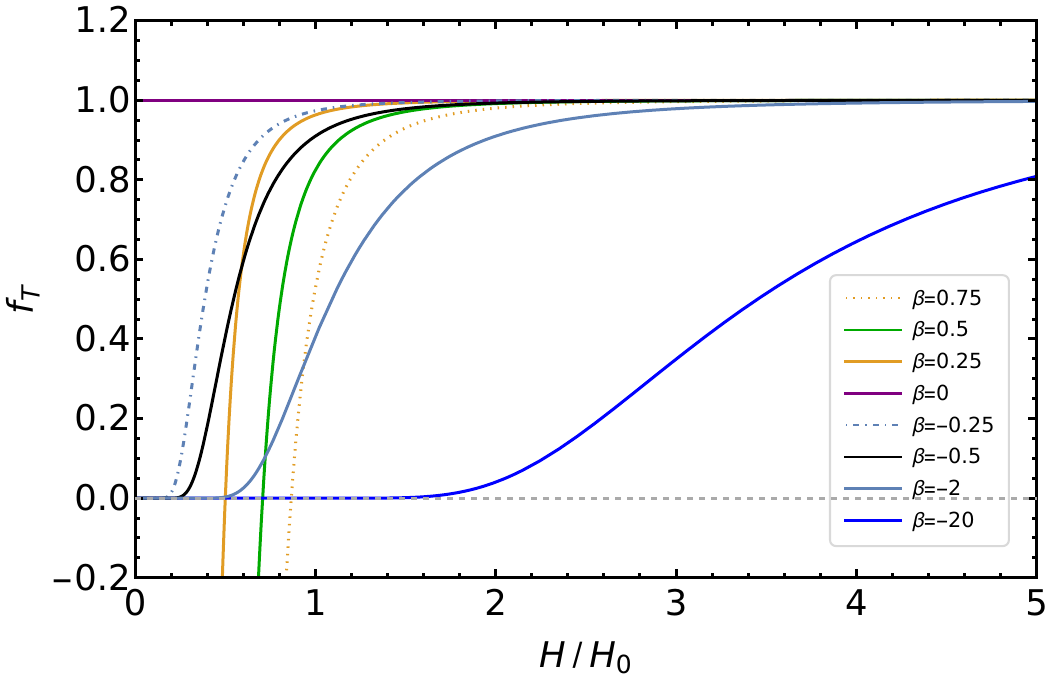}
\medskip
\hspace*{4.5mm}\includegraphics[trim =0mm  0mm 0mm 0mm, clip, width=0.465\textwidth]{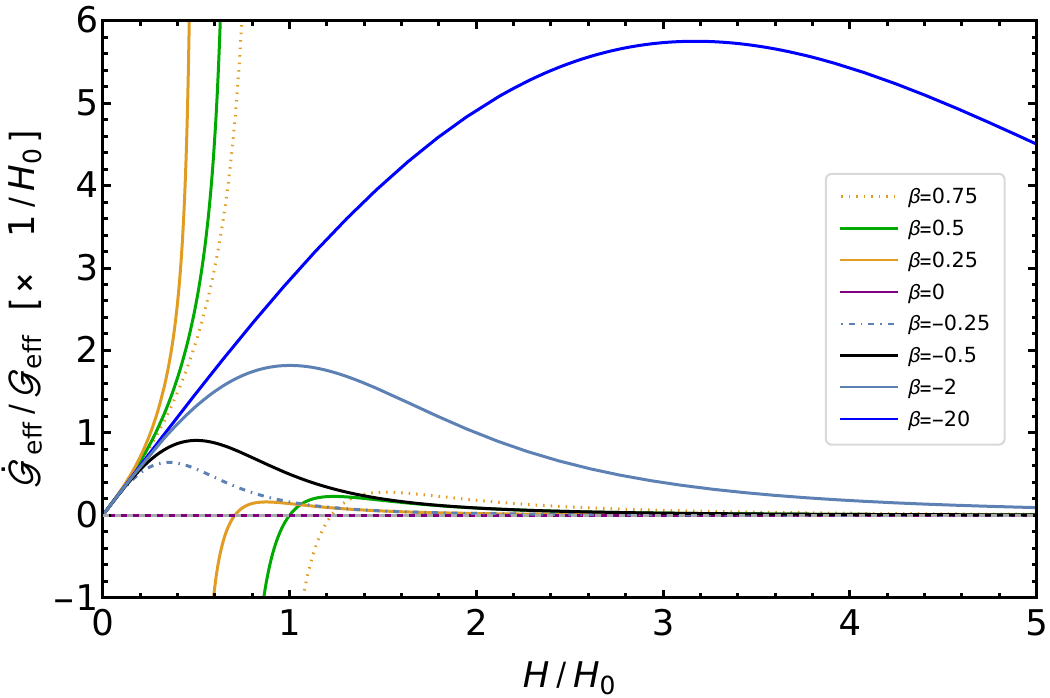} 
\end{center}
\caption{\textbf{Top panel:} $f_T$ with respect to $H/H_0$. \textbf{Bottom panel:} $\dot{\mathcal{G}}_{\rm eff}/\mathcal{G}_{\rm eff}$ scaled by $H_0$ with respect to $H/H_0$ for some chosen $\beta$ values representing regions presented in Fig~\ref{fig:evo_beta_2}.}
\label{fig:fT}
\end{figure}

After giving all theoretical capabilities of the model in terms of the quantity $f_T$, we analyse the above regions considering the positivity of $f_T$, which is required for several reasons. Probably the strongest of all is that otherwise, viz., for $f_T<0$, the gravitational waves become ghost degrees of freedom, leading to the decay of the vacuum into standard particles and ghost particles (see, e.g., Ref.~\cite{Golovnev:2018wbh}). On top of this constraint, we also need to ensure that baryonic matter follows the usual laws of gravity. If $f_T$ becomes negative, then baryons---whose energy density is positive, being made of particles/quanta with positive mass (energy)---would lead to ``antigravity,” a phenomenon that should not occur after recombination. Therefore, based on the above discussion regarding the signature of $f_T$, only Region~I is problematic in this sense. One might observe that in cases where $f_T < 0$ and $\Omega_{\rm m0} < 0$, Eq.~\eqref{eq:matpert} is still mathematically compatible with a growth of perturbations. However, it is crucial to note that when $f_T < 0$, both the matter perturbations---since $\rho_{\rm m}=\mu_0 n<0$ and $\partial \rho_{\rm m}/\partial n<0$ (or $\mu_0<0$, given that ${\rm sign}(\Omega_{\rm m0})={\rm sign}(\mu_0)$) with $n>0$ being the particle number density---and gravitational waves always become ghost degrees of freedom (the latter, regardless of the value of $\Omega_{\rm m0}$). Therefore, such a scenario cannot be considered physically acceptable. The fractional change of the effective gravitational constant, $\dot{\mathcal{G}}_{\rm eff}/\mathcal{G}_{\rm eff}=-\dot{f}_T/f_T$, modifies the  propagation of gravitational waves, governed by~\cite{Golovnev:2018wbh} 
\begin{align}
\ddot{h}_{ij}+\left(3H+ \frac{\dot{f}_{T}}{f_T}\right)\dot{h}_{ij}+\frac{k^2}{a^2}h_{ij}=0,
\end{align}
where the modification on Hubble friction term is too slight to be constrained by the currently existing gravitational wave data, as suggested by models considered in the literature~\cite{Bamba:2013ooa, Cai:2018rzd}.  In the exponential infrared teleparallel model, this term takes the form
\begin{equation}
\begin{aligned}
\frac{\dot{\mathcal{G}}_{\rm eff}}{\mathcal{G}_{\rm eff}}
=&-\frac{f_{TT}}{f_T} \dot{T} 
=-\frac{2(\beta H_0^2/{H}^2)^2}{1-\beta H_0^2/H^2}\frac{{\rm d} H}{{\rm d}\mathcal{N}},
\end{aligned}
\end{equation}
which shares the same signature as ${\rm d} H/{\rm d}\mathcal{N}$ given in Eq.~\eqref{eq:dH_dN}, except in Region~I. Figure~\ref{fig:fT} illustrates the behaviors of $f_T$ in the top panel and $\dot{\mathcal{G}}_{\rm eff}/\mathcal{G}_{\rm eff}$, scaled by $H_0$, in the bottom panel for the same $\beta$ values used in Fig.~\ref{fig:evo_H_dec}.

We note that the effective gravitational constant $\mathcal{G}_{\rm eff} $, as experienced by cosmological perturbations and defined in Eq.~\eqref{Gcosmo}, does not necessarily coincide with the effective Newtonian constant $ G_{\rm eff} $ measured within gravitationally bound/virialized structures, such as galaxies and the Solar System, where $G_{\rm eff} \approx G_{\rm N}$ must hold. In $ f(R) $ gravity theories, the Ricci scalar $R$ depends on the local environment due to the source induced by the local matter density in the modified Einstein equations, leading to variations in gravitational interactions across different settings. Analogously, in $ f(T) $ gravity, the torsion scalar $T$ is also expected to exhibit environmental dependence. Specifically, in regions where the matter density $\rho_{\rm env} \gg \rho_{\rm cosmo}(z=0) $, as one finds $R_{\rm env}\gg R(z=0)$, it is reasonable to expect $ T_{\rm env} \gg T_{\rm cosmo}(z=0) $ as well.  For instance, in a homogeneous and isotropic background, $ |T(z\gg 1)| \propto H^2(z\gg1) \simeq \rho(z\gg1)\gg \rho(z=0) $. Hence, at high redshifts, we have $f_T(z\gg1)\approx1$, leading to a GR-like limit where $\mathcal{G}_{\rm eff} \approx G_{\rm N}$, or equivalently, for $\rho(z)/\rho(z=0)\gg1$, i.e., for large energy densities compared to today's cosmological values. Therefore, we expect that the behavior in other environments (e.g., within the Solar System) is better described by the phenomenology of $f(T)$ when $T(t)\simeq T_{\rm env}$. In bound or virialized structures like galaxies and the Solar System, spacetime can be effectively considered static and decoupled from cosmological expansion, resulting in a larger torsion scalar due to the higher local densities compared to $T_{\rm cosmo}$. This indicates that the environmental dependence of $T$, governed by the local properties of spacetime, plays a crucial role in determining the behavior of $f(T)$ gravity across different physical settings.

A key requirement for recovering GR---more precisely, its teleparallel equivalent (TEGR)---in local, high-density environments is the asymptotic behavior of the function $f(T)$ and its derivative $f_T$ in the large-$T$ regime. For the $f(T)$ models satisfying the condition ${f_T|_{T\simeq T_{\rm env}\gg T_{\rm cosmo}}\approx1}$, the theory is expected to closely approximate TEGR in dense regions such as galaxies or the Solar System. In this context, a ``screening mechanism”---analogous to the chameleon mechanism~\cite{Khoury:2003aq,Khoury:2003rn} invoked in $f(R)$ gravity~\cite{Navarro:2006mw,Faulkner:2006ub,Li:2007xn,Capozziello:2007eu,Brax:2008hh,DeFelice:2010aj}---is necessary to suppress the fifth force/long-range modifications on small scales (e.g., within the Solar System) and restore TEGR, while still permitting significant deviations at cosmological scales that, for instance, can account for late-time acceleration. It should be emphasized, however, that unlike $f(R)$ gravity, the covariant formulation of $f(T)$ gravity---when recast into a conformally equivalent scalar-tensor form---does not introduce a scalar degree of freedom whose effective potential acquires an explicit dependence on the local matter density in the Einstein frame. Consequently, the standard chameleon mechanism, as formulated in scalar-tensor theories, does not directly operate in $f(T)$ gravity (see Refs.~\cite{Hohmann:2018rwf,Paliathanasis:2024sle}). Nevertheless, an \emph{effective chameleon-like screening mechanism} can still emerge in $f(T)$ theories under \textit{suitable} conditions. To clarify this, consider that in $f(R)$ gravity---even in the absence of scalar-matter couplings beyond minimal gravitational interaction---the field equations can be written as $G_{\mu\nu} + S_{\mu\nu} =\kappa \mathcal{T}_{\mu\nu}$, where $G_{\mu\nu}$ is the Einstein tensor and $S_{\mu\nu}$ encapsulates corrections from higher-order derivatives of $f(R)$, specifically involving $f_{RR}$. Taking the trace yields $-R + S = \kappa \mathcal{T}$, with $\mathcal{T}$ the trace of the energy-momentum tensor. In high-density regions where $|\kappa \mathcal{T}| \gg H_0^2$, and assuming $|S| \ll |R|$---as typically holds for viable models since $S \sim H_0^2$ in DE-dominated regimes---one recovers $|R| \sim |\kappa \mathcal{T}|$, thereby suppressing the corrections and effectively restoring GR in environments where $\mathcal{T}_{\rm env} \gg \mathcal{T}_{\rm cosmo}$. This logic closely parallels what occurs in GR with a cosmological constant: $G_{\mu\nu} + \Lambda g_{\mu\nu} =\kappa \mathcal{T}_{\mu\nu}$ with trace $-R+4\Lambda= \kappa\mathcal{T}$, on which the $\Lambda$CDM model is based. Although $\Lambda\sim H_0^2$ is essential for driving late-time cosmic acceleration on large scales, its contribution becomes negligible in high-curvature, high-density environments such as galaxies or the Solar System, where $|\mathcal{T}|\sim|R|\gg \Lambda$. In such regimes, GR without $\Lambda$, i.e., $G_{\mu\nu} \approx \kappa \mathcal{T}_{\mu\nu}$ with trace $-R= \kappa\mathcal{T}$, is effectively recovered. By analogy, in $f(T)$ gravity, if the function is constructed such that $f(T \gg H_0^2) \approx T$, then $f_T\rightarrow1$ in high-torsion regimes (high-density regions), and deviations from TEGR are suppressed. Although the mechanism differs fundamentally---being curvature-based in $f(R)$ and torsion-based in $f(T)$---the outcome is similar: modified gravity effects are negligible in dense environments, yet substantial at cosmological scales. This demonstrates that under, appropriate model choices, $f(T)$ gravity can realize an \emph{effective chameleon-like behavior} consistent with local gravity tests, e.g., solar-system tests.

Then, in the presence of such a mechanism, the Solar System constraints cannot be naively applied to these theories at cosmological scales. As a result, the effective cosmological Newtonian constant, $ \mathcal{G}_{\rm eff} $, is not directly related to the local effective Newtonian constant, denoted as $G_{\rm eff}$. This allows $ f(T) $ gravity to modify cosmological dynamics, potentially addressing phenomena such as cosmic acceleration driven by DE, while remaining consistent with observations in local environments. The post-Newtonian limit of $f(T)$ gravity has been found to be identical to that of GR~\cite{Ualikhanova:2019ygl}. It was also shown in Ref.~\cite{Chen:2014qsa} that Brans-Dicke-type nonminimal coupling of a scalar field to torsion results in the same parameterized post-Newtonian (PPN) parameters as in GR. Similarly, Ref.~\cite{Emtsova:2019qsl} demonstrated that other scalar field couplings without kinetic terms exhibit no deviation from GR's PPN parameters in the massless scalar field case, and for the massive case, only the $\gamma$ parameter was calculated, which was also found to agree with GR. Therefore, these findings might be considered as an indication that $f(T)$ gravity is endowed with a chameleon-like mechanism similar to that of $f(R)$~\cite{Navarro:2006mw,Faulkner:2006ub,Li:2007xn,Capozziello:2007eu,Brax:2008hh}. Nonetheless, given Eq.~\eqref{eq:matpert}, the cosmological dynamics of matter perturbations $ \delta_{\rm m} $ will impact observables related to the growth of large-scale structures. These could provide constraints from several experiments and may even help to alleviate the $S_8$ tension due to the deviation of the cosmological Newtonian constant $\mathcal{G}_{\rm eff}$ from $G_{\rm N}$. Furthermore, even if the chameleon mechanism is at work at astrophysical scales, as soon as a gravitational wave feels the cosmological environment during its propagation---for instance, a wave produced in a galaxy a few (say, $10-100$) Mpc away from us---its propagation will also reflect the cosmological behavior. Future observational tests, including measurements of the growth rate of cosmic structures, redshift-space distortions, and weak lensing surveys, can provide stringent constraints on $ f(T) $ gravity. Exploring these phenomenological implications is beyond the scope of the current work and will be left as a future project.

\section{Exploring viable cosmologies}  
\label{sec:cosm}

In this section, we investigate a compelling theoretical feature of the $f(T)$ model under consideration: the possibility that the effective DE density can attain negative values at high redshifts, independent of the presence of a negative bare cosmological constant. Typically, a self-gravitating fluid component with $\rho < 0$ in the background would be regarded as problematic. For a simple matter component, the energy density $\rho_{\phi}$ remains positive unless a ghost-like degree of freedom is present. In standard quintessence models, if a negative bare cosmological constant is included, the total energy density of the quintessence field and the cosmological constant could indeed become negative, provided the cosmological constant is sufficiently large and negative. However, in the absence of such a cosmological constant, where the quintessence potential vanishes at its absolute minimum, the energy density of the quintessence field remains strictly non-negative.

In the context of $f(T)$ gravity, as we will demonstrate in the following discussion, the theory can enter an era where $\rho_{\rm T} < 0$ even without the need for a cosmological constant.\footnote{For instance, at early times, we find $\lim_{H/H_0 \to \infty} \rho_{\rm T} = 3\beta H_0^2/\kappa < 0$ for $\beta < 0$, while for any finite value of $H$, $\rho_{\rm T} + p_{\rm T} \neq 0$. This shows that the $f(T)$ modification does not act as a cosmological constant.
By construction, we assume that at the present time, $\kappa \rho_{\rm T}(z=0)/(3H_0^2) = 1 - \Omega_{\rm m0}$, which is positive for $\Omega_{\rm m0} < 1$.} Moreover, the theory remains free from ghost instabilities, as $f_{T} > 0$ is preserved throughout. This intriguing feature may lead to nontrivial phenomenological consequences. While a cosmological constant can be introduced into the $f(T)$ framework to enrich its phenomenological scope, our analysis in this section is limited to the scenario without a cosmological constant. In Sec.~\ref{sec:lambda}, we will expand the discussion to explore viable cosmologies by incorporating $\Lambda$ into the action \eqref{eq:action}, while retaining the function~\eqref{eq:model} describing the exponential infrared teleparallel model.

\subsection{Effective DE interpretation} 
To investigate the exponential infrared model from the perspective of dynamical DE models, we treat the additional geometrical terms arising from the $f(T)$ modification in the modified Friedmann equations as an effective DE component, see Eqs.~\eqref{eq:rho}-\eqref{eq:pT_gen}. Note that the $T$ part of the action~\eqref{eq:action} generates the standard $3 H^2$ term, and thereby, the extra terms stem from the variation of the $f(T) - T$ contribution. Similar to Eq.~\eqref{redshift}, we can express all quantities in terms of the Hubble parameter $H$, thus allowing their $z$-dependencies to be written in parametric form. To proceed, we rewrite the Friedmann equation from Eq.~\eqref{eq:fried1} as
\begin{align}  \label{rho_darkenergy}
3H^{2}&= \kappa \rho_{\rm m0} (1+z)^3+ \kappa \rho_{\rm T}, 
\end{align}
where the energy density of the torsional DE can be extracted as
\begin{align}
\label{rho}
    \rho_{\rm T}(H)=&\frac{3 H^2}{\kappa}\left[1-\left(1-2\beta H_0^2/H^2\right)e^{\beta H_0^2/H^2}\right],
\end{align}
which can also be obtained by substituting Eq.~\eqref{eq:model} along with the relation~\eqref{eq:THreln} into Eq.~\eqref{eq:rhoT_gen}. Using Eq.~\eqref{eq:p}, or equivalently, the fact that the torsional DE satisfies the local energy-momentum conservation law, viz., $\dot{\rho}_{\rm T}+3H(\rho_{\rm T}+p_{\rm T})=0$, the pressure of the torsional DE is given by
 \begin{align}
p_{\rm T}(H)=-\frac{3\beta H_0^2}{\kappa}\left(\frac{1+2\beta H_0^2/H^2}{1-\beta H_0^2/H^2+2\beta^2H_0^4/H^4}\right).
\label{pres}
\end{align}
Consequently, the corresponding equation of state (EoS) parameter of the torsional DE reads
\begin{equation}
\begin{aligned} \label{eos}
w_{\rm T}=-&\frac{\beta H_0^2/H^2}{1-\beta H_0^2/H^2+2\beta^2H_0^4/H^4}\\
&\times \frac{1+2\beta H_0^2/H^2}
{1-\left(1-2\beta H_0^2/H^2\right)e^{\beta H_0^2/H^2}}. 
\end{aligned}
\end{equation}

To conduct a reasonable quantitative analysis of the effective DE model, whose theoretical basis is outlined above, we now present a preliminary forecast of the relevant cosmological parameters. We fix the angular scale of the sound horizon, $\theta_*=r_*/D_{M}(z_*)$, and the present-day physical matter density, $\Omega_{\rm m0}h^2$, to the Planck CMB data constraints, assuming the base $\Lambda$CDM model, ensuring reasonable consistency with the CMB power spectra as described below. Since the modification considered in Eq.~\eqref{eq:model} becomes effective only in the post-recombination epoch, we do not expect significant deviations in the comoving sound horizon at last scattering,  $r_*=\int_{z_*}^{\infty} c_{\rm s} H^{-1} {\rm d} z$, which is determined primarily by the pre-recombination Universe, from that predicted by $\Lambda$CDM in the $f(T)$ model under consideration in our work. Here, $c_{\rm s}$ is the sound speed in the plasma and $z_* \approx 1090$ is the redshift of last scattering. Given that $\theta_*$ is measured in a nearly model-independent manner with a precision of  0.03\%~\cite{Planck:2018vyg}, we equivalently fix the comoving angular diameter distance to last scattering, given by $D_{M}(z_*)=\int_{0}^{z_*} c H^{-1} {\rm d} z$. Accordingly, we adopt the value from the Planck (TT,TE,EE+lowE+lensing) best-fit values for the base $\Lambda$CDM model~\cite{Planck:2018vyg}; $D_{M}(z_*)= 13869.6 {\;\rm Mpc}$, tightly constrained through the measurement of $\theta_*$ ($100\theta_*=1.041085$), along with the CMB-based constraint $\Omega_{\rm m0}h^2=0.14314$, which implies an inverse correlation between $\Omega_{\rm m0}$ and $H_0$. Here, $h=H_0/(100\,{\rm km\,s}^{-1}{\rm Mpc}^{-1})$ represents the dimensionless reduced Hubble constant. The Planck (TT,TE,EE+lowE+lensing) best-fit value for the Hubble constant is $H_0 = 67.32 \,{\rm km\,s}^{-1}{\rm Mpc}^{-1}$ for the $\Lambda$CDM model. Using our method, we obtain $H_0 = 67.23 \,{\rm km\,s}^{-1}{\rm Mpc}^{-1}$, confirming the robustness of our estimations. However, when we exclude radiation from our calculations, our estimation deviates significantly, yielding $H_0 = 68.95 \,{\rm km\,s}^{-1}{\rm Mpc}^{-1}$ for $\Lambda$CDM. This discrepancy arises because, although radiation can be neglected in the late Universe, it still affects $H(z)$ at high redshifts, particularly around recombination, when the radiation energy density cannot be entirely ignored. To compensate for this energy density deficiency, we use a slightly larger value for the physical matter density, $\Omega_{\rm m0}h^2 = 0.1444$, to improve our estimations. With this adjustment, excluding radiation, we find $H_0 = 67.70 \,{\rm km\,s}^{-1}{\rm Mpc}^{-1}$---an estimation accurate to within 1\%, which is sufficient for the purposes of our preliminary investigations. In the $f(T)$ model for phantom torsional DE (i.e., for the positive exponent case achieved with $\beta_+$), this approach yields $H_0 = 72.36 \,{\rm km\,s}^{-1}{\rm Mpc}^{-1}$, which matches the value obtained from robust observational analysis, $H_0 = 72.24 \pm 0.64 \,{\rm km\,s}^{-1}{\rm Mpc}^{-1}$, as reported in Ref.~\cite{Hashim:2021pkq}. Without compensating for the missing radiation component, we would obtain $H_0 = 73.79 \,{\rm km\,s}^{-1}{\rm Mpc}^{-1}$, a value that deviates considerably from the constraints given in Ref.~\cite{Hashim:2021pkq}. Therefore, to conduct a fair and accurate comparison between the models (studied here theoretically without including radiation for simplicity), we replace $\Omega_{\rm m0}h^2 = 0.14314$ with $\Omega_{\rm m0}h^2 = 0.1444$ to use along with the Planck best-fit value for $D_{M}(z_*) = 13869.6 \,{\rm Mpc}$. This adjustment ensures that our results are consistent with the Planck-CMB constraints on the $\Lambda$CDM and $f(T)$ models for phantom torsional DE ($\beta_+$), as present in the literature, with an accuracy of 1\%. Additionally, this ensures that our estimations of the free parameters of the models and the plots provided throughout the paper are not arbitrary but are consistent with Planck CMB data at a reasonable level, even though we do not include radiation.

Consequently, utilizing our method, we obtain $H_0=72.36 \,{\rm km\,s}^{-1}{\rm Mpc}^{-1}$ and $\Omega_{\rm m0}=0.276$ ($\beta_{+} = 0.408$) for Region~II, and $H_0=84.54 \,{\rm km\,s}^{-1}{\rm Mpc}^{-1}$ and $\Omega_{\rm m0}=0.202$ ($\beta_{-} = -3.736$) for Region~IV.\footnote{\label{FN:7}Given that the two models are expected to exhibit identical pre-recombination dynamics to remain consistent with CMB observations, one may also adopt the $68\%$ confidence level constraints from \textit{Planck} 2018 for the base $\Lambda$CDM model: $D_M(z_*) = 13873 \pm 25\;{\rm Mpc}$ and $\Omega_{\rm m0} h^2 = 0.1430 \pm 0.0011$~\cite{Planck:2018vyg}. Under these conditions, the resulting deviations in $\beta$ and $H_0$ from their best-fit values are at most at the level of $2\%\text{--}3\%$. Accordingly, the parameter values we adopt can be regarded as a representative subset of viable choices, and such slight variations do not affect the qualitative conclusions of our analysis. Nonetheless, a more comprehensive observational analysis is required to derive robust parameter constraints; a dedicated follow-up study is currently in preparation.} For comparison, applying the same constraints to the $\Lambda$CDM model yields $H_0=67.70 \,{\rm km\,s}^{-1}{\rm Mpc}^{-1}$ and $\Omega_{\rm m0}=0.315 $. The $\beta_{+}$ case has been well studied in the literature, where it produces a torsional DE with phantom behavior~\cite{Awad:2017yod,Hashim:2020sez,Hashim:2021pkq}.  In the following sections, we examine both cases in comparison with the $\Lambda$CDM model, presenting illustrative figures. First, we revisit the $\beta_{+}$ model in Sec.~\ref{sec:phantom}, followed by an exploration of the $\beta_{-}$ model in Sec.~\ref{sec:signchange}, which effectively generates a DE with zero-crossing energy density---a scenario excluded in previous studies~\cite{Awad:2017yod,Hashim:2020sez,Hashim:2021pkq}.

\begin{figure*}[t!]
\par
\begin{center}
\includegraphics[trim =0mm  0mm 0mm 0mm, clip, width=0.48\textwidth]{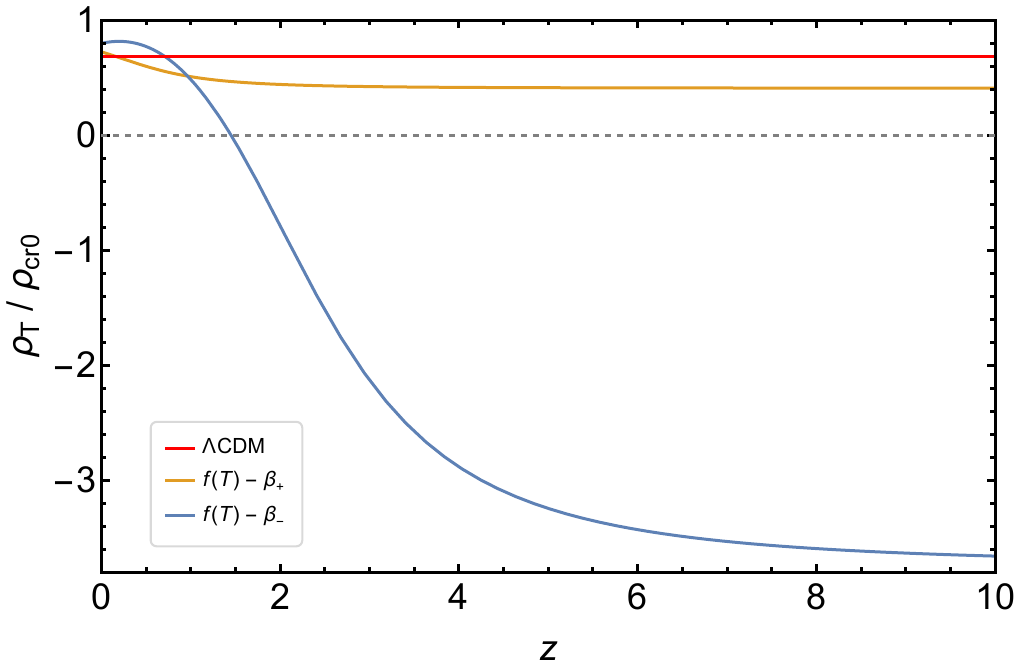}
\includegraphics[trim =0mm  0mm 0mm 0mm, clip, width=0.48\textwidth]{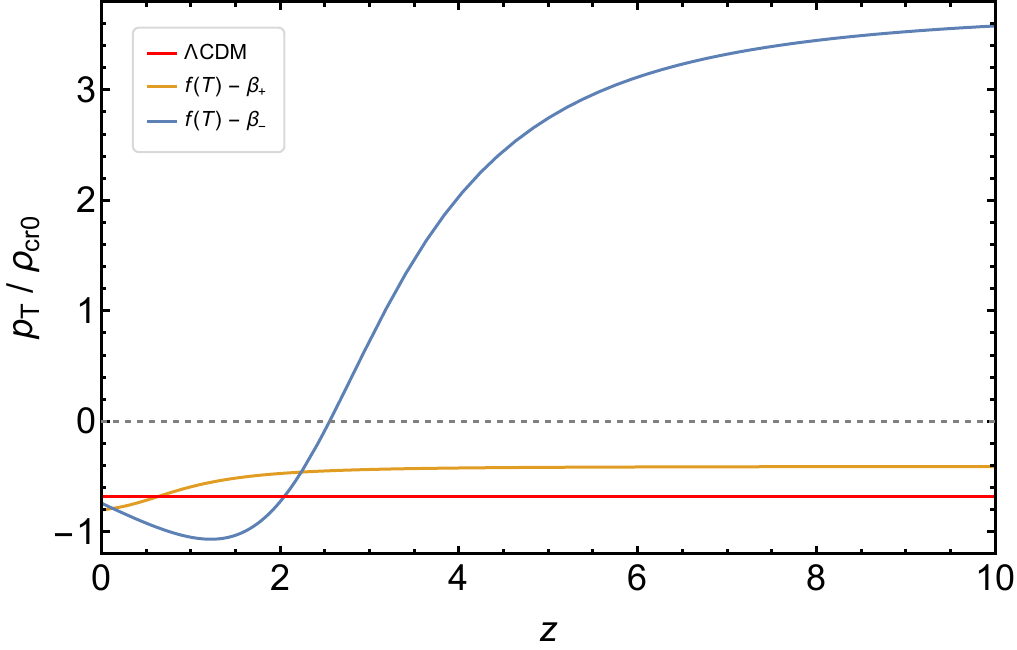} 
\hspace*{-1.5mm}\includegraphics[trim =0mm  0mm 0mm 0mm, clip, width=0.48\textwidth]{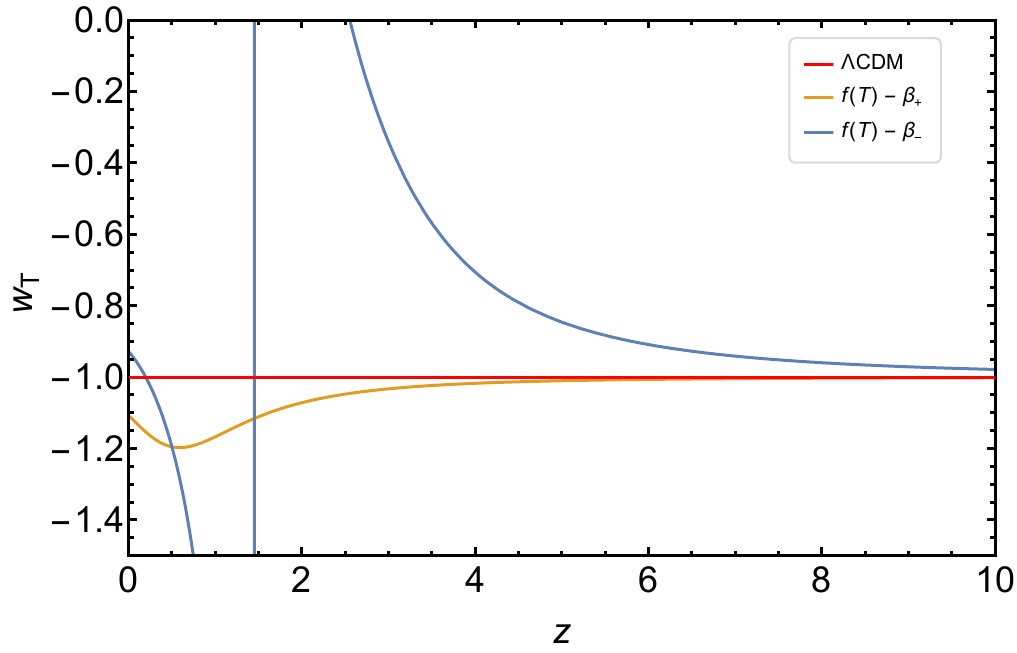}
\includegraphics[trim = 0mm  0mm 0mm 0mm, clip, width=0.48\textwidth]{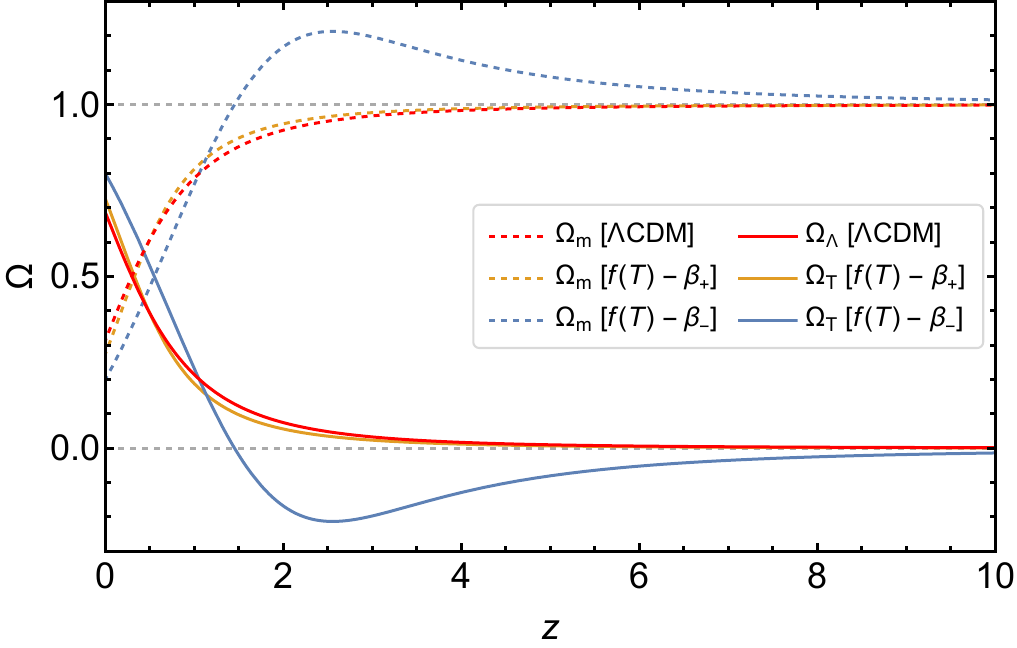}
\end{center}
\caption{\textbf{Top left panel:}  $\rho_{\rm T}(z)/\rho_{\rm cr0}$ (energy density of torsional DE scaled by present-day critical density). \textbf{Top right panel}: $p_{\rm T}(z)/\rho_{\rm cr0}$ (pressure of torsional DE scaled by present-day critical density). \textbf{Bottom left panel}:  $w_{\rm T}(z)$ (EoS parameter of torsional DE). \textbf{Bottom right panel}: $\Omega(z)$ (density parameters) of matter (dashed curves) and torsional DE (solid curves). To ensure the consistency with the CMB data, we use $H_0=67.70  \,{\rm km\,s}^{-1}{\rm Mpc}^{-1}$ for the $\Lambda$CDM model shown by the red curve, $H_0=72.36  \,{\rm km\,s}^{-1}{\rm Mpc}^{-1}$ for the phantom DE model ($\beta_{+}$) shown by the orange curve, and $H_0=84.54  \,{\rm km\,s}^{-1}{\rm Mpc}^{-1}$  for the sign-changing DE model ($\beta_{-}$) shown by the blue curve. 
}
\label{fig:wde,deltawde}
\end{figure*}

\subsection{$\beta_{+}$ case: Phantom torsional DE}
\label{sec:phantom}  
We now explore some features of the phantom DE model achieved with $\beta_{+}$---remaining consistent with the color used to represent Region~II, the solid orange curve, in Fig.~\ref{fig:evo_beta_2}---using a series of illustrative plots. The top panels of Fig.~\ref{fig:wde,deltawde} depict the evolution of the energy density, $\rho_{\rm T}(z)$, and the pressure, $p_{\rm T}(z)$, of the torsional DE, both scaled by $\rho_{\rm cr0}$. The bottom left panel presents the corresponding EoS parameter, $w_{\rm T}(z)$, while the bottom right panel shows the evolution of the density parameters for matter and torsional DE, comparing the phantom model with $\beta_{+}$ (orange curves) to the $\Lambda$CDM model (red curves).

At large redshift values ($z \gtrsim 5$), the torsional DE behaves like a cosmological constant, with its EoS parameter becoming indistinguishably close to $-1$, specifically, $w_{\rm T}(z)\to-1$ as $z\to\infty$. Given the lack of significant deviation from $-1$ in $w_{\rm T}(z)$, the evolution of the torsional DE density parameter $\Omega_{\rm T}(z)$ closely mirrors that of $\Omega_{\Lambda}(z)$ in the $\Lambda$CDM model. While the present-day energy density of torsional DE is $0.724 \rho_{\rm cr0}$, it decreases monotonically with increasing $z$, approaching asymptotically $0.408 \rho_{\rm cr0}$, as the ratio $\rho_{\rm T}/\rho_{\rm cr0}\to\beta$ as $z\to\infty$.  The EoS parameter $w_{\rm T}(z)$ stays indistinguishably close to $-1$ for $z>5$, but becomes noticeably less than $-1$ for $z<5$, with the maximum deviation from $-1$ occurring around $\sim0.5$, where $w_{\rm T}$ reaches $-1.2$. Nevertheless, the EoS parameter remains within observationally acceptable bounds. As shown in the bottom right panel of Fig.~\ref{fig:wde,deltawde}, the matter (dashed curve) and torsional DE (solid curve) have present-day values of  $\Omega_{\rm m0}=0.276 $ and $\Omega_{\rm T0}=0.724$, respectively. At $z=0$, the torsional DE still exhibits a significant phantom character  with an EoS parameter value of $w_{\rm{T}}(z=0)=-1.11$. Importantly, $\rho_{\rm T}$ remains positive throughout the Universe's evolution in this case. As seen in the bottom panel of Fig.~\ref{fig:H-and-q}, for large redshifts ($z\gg1$) the model recovers the deceleration parameter of the matter-dominated universe, similar to $\Lambda$CDM, with $q=0.5$. The onset of accelerated expansion occurs slightly earlier in this model at $z_{\rm{tr}} \approx 0.73$, compared to $z_{\rm{tr}} \approx 0.63$ in $\Lambda$CDM. The present-day deceleration parameter is $q_0 = -0.70$, whereas it is $-0.53$ in $\Lambda$CDM. In the far future limit ($z \to -1$), we have $q \to -1$ as $H \to 65.39 \,{\rm km\,s}^{-1}{\rm Mpc}^{-1}$, resulting in a de Sitter universe. In the top panel of Fig.~\ref{fig:H-and-q}, we plot the comoving Hubble parameter (viz., the expansion rate $\dot{a}=H(z)/(1+z)$) as a function of redshift. It can be seen that the phantom nature of the effective DE introduces a late-time deviation in $H(z)$ relative to $\Lambda$CDM. Specifically, $H(z)$ is lower than $H_{\Lambda \rm CDM}(z)$ for $z \gtrsim 0.5$, but becomes higher than $H_{\Lambda \rm CDM}(z)$ for $z \lesssim 0.5$, ensuring that $D_M(z_*)$ remains consistent with its value obtained from \textit{Planck}-$\Lambda$CDM. Consequently, this deviation implies an increased value of $H_0$ compared to the \textit{Planck}-$\Lambda$CDM prediction. In particular, the enhanced late-time acceleration, due to the pronounced phantom character of the torsional DE at low redshifts, leads to a higher value of $H_0$, with the model predicting $H_0=72.36 \,{\rm km\,s}^{-1}{\rm Mpc}^{-1}$, in excellent agreement with the Supernovae and $H_0$ for the Equation of State of Dark Energy (SH0ES) $H_0$ measurement of $73.04  \pm 1.04 \,{\rm km\,s}^{-1}{\rm Mpc}^{-1}$~\cite{Riess:2021jrx}. Additionally, a visual comparison suggests that the model describes the BAO data at high redshifts better than the $\Lambda$CDM model.\footnote{For illustrative purposes, we include in~\cref{fig:H-and-q}, as well as in~\cref{fig:dec-comoving} of Sec.~\ref{sec:lambda}, the green error bars corresponding to the  pre-DESI BAO data, viz., the completed extended Baryon Oscillation Spectroscopic Survey (eBOSS) BAO data~\cite{eBOSS:2020yzd}. A comprehensive observational analysis of the model—based on the latest datasets, including the recent DESI Data Release (DR) 2 BAO measurements~\cite{DESI:2025zgx}—is currently underway, as noted in footnote~\ref{FN:7}, and will be presented in a follow-up study.}

\subsection{$\beta_{-}$ case: Sign-changing torsional DE}
\label{sec:signchange}
We now explore some features of the sign-changing DE model achieved with  $\beta_{-}$, remaining consistent with the color used to represent Region IV, the solid blue curve, in Fig.~\ref{fig:evo_beta_2}. As seen in the top left panel of Fig.~\ref{fig:wde,deltawde}, the energy density of phantom DE (orange curve), which is achieved for $\beta_{+}$, decreases with increasing redshift but remains positive throughout. In contrast, the energy density of torsional DE with $\beta_{-}$ (blue curve) takes negative values at high redshifts (in the past), demonstrating a unique sign-changing behavior. 
As a concrete example, in this study, with the choice of $\beta_{-}$, we show that torsional DE provides a transition from a negative DE in the early Universe to a positive one in the late Universe and in this regard, the cosmological models deserve to be investigated for this choice as well.

The energy density of torsional DE passes from zero at a specific redshift, denoted by $z_{\dagger}$, as follows: 
\begin{align} 
\label{eq:rho0}
\rho_{\rm T}(z=z_{\dagger})=0, 
\end{align} 
from which Eq.~\eqref{rho_darkenergy} is reduced to that of a matter-dominated FLRW universe:  $H_{\dagger}^2/H_0^2=\Omega_{\rm m0}(1+z_{\dagger})^3$ where $H_{\dagger} \equiv H(z=z_{\dagger})$, then, this allows us to eliminate the Hubble parameter from Eq.~\eqref{rho} with Eq.~\eqref{eq:rho0} and express the redshift as
\begin{align} \label{z-sign}
z_{\dagger}=\left\{\frac{\Omega_{\rm m0}}{\beta_{-}} \left[\frac{1}{2}+W_{-1}\left(-\frac{1}{2\sqrt{e}}\right)\right]\right\}^{-1/3}-1,
\end{align}
where $W_{-1}$ is the $k=-1$ branch of the Lambert-$W$ function, $W_{k}$, and $\Omega_{\rm m0}$ is determined by $\beta_{-}$ through the constraint equation~\eqref{eq:constraint}. At $z_{\dagger}$, the energy density of torsional DE changes sign, causing a singularity (pole) in the EoS parameter, as shown in the bottom left panel of Fig.~\ref{fig:wde,deltawde}. From Eq.~\eqref{eos}, we find that ${\lim_{H\to H_\dagger^\pm}w_{\rm T}(H)=\pm\infty}$, which physically implies that beyond $z_{\dagger}$ in the past, as the sign-changing torsional DE takes negative values, the EoS parameter increases from $w_{\rm T}(z\gg z_{\dagger})\approx -1$ to positive infinity. The singularity (pole) occurs at $z_{\dagger}$ when the torsional DE density crosses zero, from negative to positive.  After this transition, the EoS parameter increases from negative infinity in the phantom region, approaching a finite value above the PDL, i.e., in the quintessence region, yet remaining close to $-1$, as $z$ decreases at low redshifts. As shown in Ref.~\cite{Ozulker:2022slu} (see also Refs.~\cite{Akarsu:2019hmw,Adil:2023exv}), such a singularity (pole),  ${\lim_{z\to z_\dagger^\pm}w_{\rm T}(z)=\pm\infty}$ (and not the other way around), is necessary for a minimally interacting source---in this case DE---whose energy density changes sign and becomes positive in the late Universe.

\begin{figure}[t!]
\par
\begin{center}
\hspace*{0mm}\includegraphics[trim = 0mm  0mm 0mm 0mm, clip, width=0.48\textwidth]{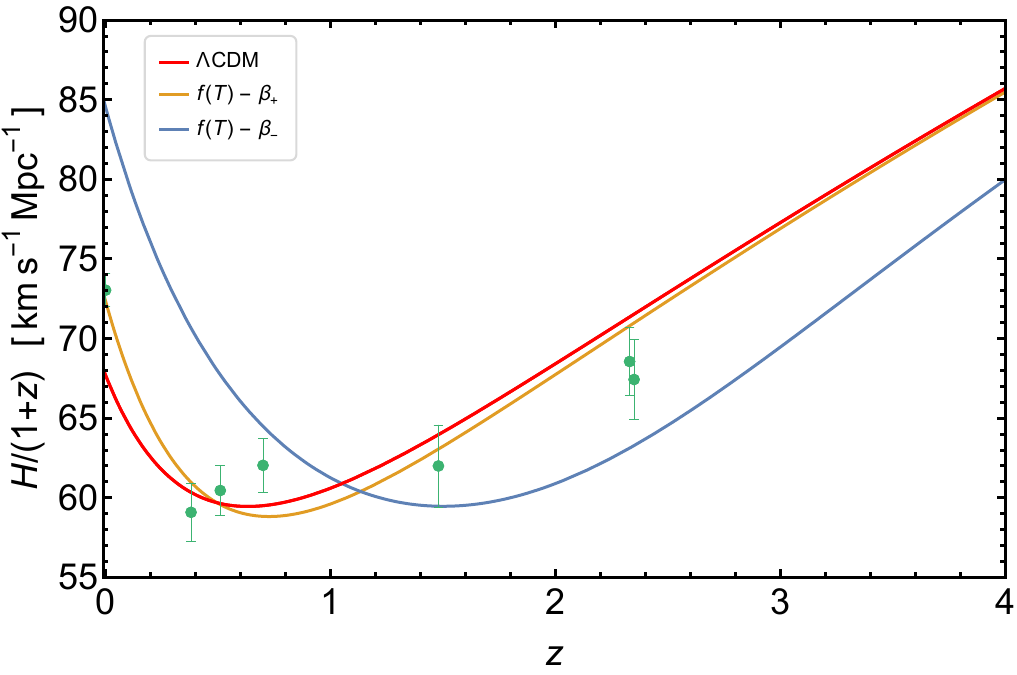} 
\hspace*{-2.mm}\includegraphics[trim = 0mm  0mm 0mm 0mm, clip, width=0.497\textwidth]{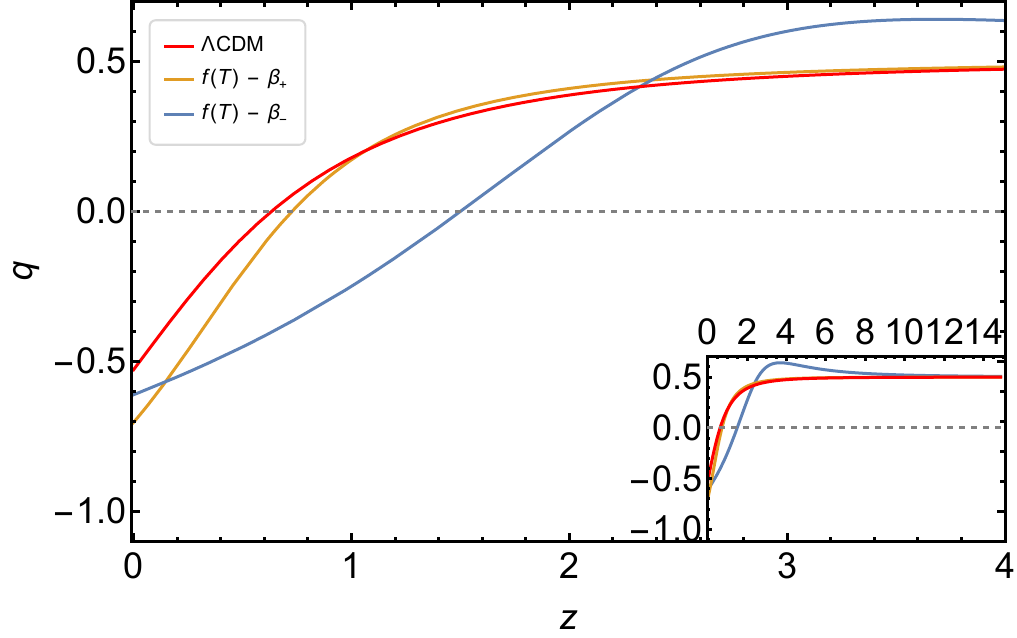}
\end{center}
\caption{\textbf{Top panel:} $\dot{a}=H(z)/(1+z)$ (comoving Hubble parameter). \textbf{Bottom panel:} $q(z)$ (deceleration parameter). The green bars corresponds to SH0ES collaboration measurement~\cite{Riess:2021jrx} and clustering measurements for the BAO samples in Ref.~\cite{eBOSS:2020yzd}; BOSS DR12 consensus Galaxy (from $z_{\rm{eff}} = 0.38, 0.51$), eBOSS DR16 LRG (from $z_{\rm{eff}}= 0.70$), eBOSS DR16 Quasar (from $z_{\rm{eff}}= 1.48$), eBOSS DR16 Lyman$\alpha$ (Ly$\alpha$)-Ly$\alpha$ (from $z_{\rm{eff}} = 2.33$) and eBOSS DR16 Ly$\alpha$-quasar (from $z_{\rm{eff}} = 2.33$ but shifted to $z_{\rm{eff}} = 2.35$ in the figures for visual clarity) measurements. } 
\label{fig:H-and-q}
\end{figure}

While in the case of $\beta_{+}$ the torsional DE remains strictly in the phantom region, in the case of $\beta_{-}$ the torsional DE exhibits quintessence behavior in two distinct periods: a long period before the sign transition and the present-day neighborhood after the transition, while displaying phantom behavior during the intervening period following the transition, specifically at $0.2 \lesssim z < z_{\dagger}$. However, its phantom character is much more pronounced compared to the case of $\beta_{+}$, as its EoS exhibits significantly large negative values during this epoch, beginning from negative infinity just after the transition at $z_{\dagger}=1.45$, as calculated from Eq.~\eqref{z-sign} using the relevant parameters. Note that the torsional DE yields an EoS parameter in the quintessence region in the pre-transition era ($z > z_{\dagger}$), asymptotically approaching $-1$ with increasing $z$, but with a negative energy density, which reaches a finite  value, settling into a plateau at $ -3.736  \rho_{\rm cr0}$, as can be seen from its behavior for $z \gtrsim 8$---consistent with the relation $\rho_{\rm T}/\rho_{\rm cr0} \to \beta$ as $z \to \infty$. In contrast, at low redshifts, for $z \lesssim 0.2$, the torsional DE returns to the quintessence region, yet this time with positive energy density. Specifically, we find $w_{\rm T}(z=0) = -0.93$ with the present-day energy density $0.798\rho_{\rm cr0}$. The bottom right panel of Fig.~\ref{fig:wde,deltawde} shows the density parameters, whose present-day values are $\Omega_{\rm m0}=0.202$ for matter (dashed curve) and $\Omega_{\rm T0}=0.798$ for torsional DE (solid curve). We note that $\Omega_{\rm{m}}(z)>1$ in the pre-transition era due to the negative values of the torsional DE in this epoch. Nevertheless, $\Omega_{\rm{m}}(z)$ decreases monotonically for $z \gtrsim 2.5$, and at large redshifts, say for $z \gtrsim 8$, we find $\Omega_{\rm{m}}(z) \approx 1$, implying that $\Omega_{\rm{T}}(z) \approx 0$. This indicates a recovery of the matter-dominated universe, similar to the behavior in $\Lambda$CDM. For $z \gtrsim 8$, much like the $\beta_{+}$ case, the $\beta_{-}$ case also becomes indistinguishable from the standard $\Lambda$CDM model. As seen from the bottom panel of Fig.~\ref{fig:H-and-q}, the onset of accelerated expansion occurs quite earlier in this model at $z_{\rm{tr}} \approx 1.50$---much different than $z_{\rm{tr}} \approx 0.63$ in $\Lambda$CDM and $z_{\rm{tr}} \approx 0.73$ in phantom model---as a consequence of the torsional DE taking large negative energy density values in the past. The deceleration parameter takes the value $q_0 = -0.61$ today (compared to $q_0 = -0.53$ in $\Lambda$CDM), while it reaches values larger than $q = 0.5$ at $z = 2.6$, with a maximum of $q \approx 0.64$ at $z \approx 3.7$ due to the negative torsional DE density, and then decreases with increasing redshift, ultimately settling into the usual matter-dominated era, as in $\Lambda$CDM ($q = 0.5$), for $z \gtrsim 10$.  Specifically, $q > 0.5$ at redshifts higher than $z \approx 2.5$, where the energy density parameter of torsional DE also approaches its minimum values, up until $z \approx 10$.  In the far future limit ($z \to -1$), we have $q \to -1$ as $H \to 0$.  Hence, the Universe was initially sourced by an AdS-like torsional DE, transitions to a dS-like one today, and in the far future arrives at a Minkowskian fixed point (see Sec.~\ref{sec:6p} for details). Due to the sign-changing nature of the effective DE, deviations in $H(z)$ relative to $\Lambda$CDM begin at higher redshifts compared to the phantom model. Specifically, $H(z)$ is significantly lower than $H_{\Lambda \rm CDM}(z)$ for $z \gtrsim 1$, but then becomes much higher than $H_{\Lambda \rm CDM}(z)$ for $z \lesssim 1$, ensuring that $D_M(z_*)$ remains consistent with the value obtained from \textit{Planck}-$\Lambda$CDM. Consequently, this deviation implies an over enhanced value of $H_0$ compared to the \textit{Planck}-$\Lambda$CDM prediction. In particular, the transition of the torsional DE from large negative to positive values at low redshifts leads to a higher value of $H_0$, with the model predicting $H_0=84.54 \,{\rm km\,s}^{-1}{\rm Mpc}^{-1}$, indicating an augmented accelerated expansion of the Universe. It is also worth noting that, correlated with the enhancement in the Hubble constant, the sign-changing DE model conflicts with the BAO Ly$\alpha$ data in a different way compared to $\Lambda$CDM, preferring lower $H(z)$ values than the lower limits at $z>z_{\dagger}$.

The insight gained from this section is that the particular $f(T)$ model under consideration offers a promising mechanism to alleviate major cosmological tensions. Our findings highlight the potential of $f(T)$ gravity in addressing these tensions and call for further investigation, either by exploring new functionals or re-examining existing models, as we have done for the exponential infrared teleparallel model. The possibility of effective DE densities assuming negative values in the past might have been overlooked in previous studies, as the relevance of such dynamics has only recently been recognized in connection with cosmological tensions, despite $f(T)$ gravity being studied long before these tensions became a central discussion in cosmology.

\section{Inclusion of cosmological constant $\Lambda$: Extended viable cosmologies} \label{sec:lambda}

In this section, we extend the exponential infrared cosmological model from the previous section by incorporating a cosmological constant, modifying it as $f(T) \rightarrow f(T) + 2\Lambda$ with $f(T) = Te^{\beta T_0/T}$, where the $\beta = 0$ limit corresponds to TEGR with a cosmological constant, thereby recovering the standard $\Lambda$CDM model. We will explicitly demonstrate that, with this extension, model families consistent with the CMB power spectra and in strong agreement with the SH0ES $H_0$ measurement are broadened, leading to richer phenomenological possibilities. This extension introduces one additional free parameter compared to both the exponential infrared cosmological model and the standard $\Lambda$CDM model. We will show that even this simplest modification to the original model significantly expands the range of viable cosmologies.

 We generalize the $f(T)$ action given in Eq.~\eqref{eq:action} by introducing a cosmological constant ($\Lambda$),
\begin{equation}
   \mathcal{S}=\int {\rm d}^4 x\; \vert \vert e \vert \vert \left\{-\frac{1}{2\kappa} \left[f(T)+2 \Lambda\right]+\mathcal{L}_{\rm{m}}\right\},
\label{eq:actionL}
\end{equation}
from which the modified Friedmann equations~\eqref{eq:rho}-\eqref{eq:p} are extended due to the cosmological constant, as follows:
\begin{align}  
\label{eq:rhoL}
3H^2&=\kappa\rho+\kappa\rho_{\rm T}+\Lambda, \\
-2\Dot{H}-3H^2&=\kappa p+\kappa p_{\rm T}-\Lambda,
\label{eq:pL}
\end{align}
where $\rho_{\rm T}$  and $p_{\rm T}$ are as given by Eqs.~\eqref{eq:rhoT_gen} and~\eqref{eq:pT_gen}, respectively. To investigate extensions of the viable cosmologies discussed in Sec.~\ref{sec:cosm}, we retain the exponential infrared teleparallel model described by Eq.~\eqref{eq:model} along with the assumption of pressureless matter. Thus, Eqs.~\eqref{eq:fried1} and~\eqref{eq:constraint} are modified as
\begin{align}  \label{eq:FRL}
\left(\frac{H^2}{H_0^2}-2\beta \right)e^{\beta H_0^2/H^2}&= \Omega_{\rm m0}(\beta,\Lambda)\,(1+z)^3+\Omega_{\Lambda0},\\
\Omega_{\rm m0}(\beta,\Lambda)&=(1-2\beta)e^{\beta}-\Omega_{\Lambda0}, \label{constraintL}
\end{align}
where the present-day density parameter of the cosmological constant is defined as $\Omega_{\Lambda0}=\Lambda/(3H_0^2)$. We note that the field equations were derived by extending the $f(T) = Te^{\beta T_0/T}$ model with the addition of a cosmological constant, $\Lambda$; however, they can also be equivalently interpreted as a one-parameter exponential infrared teleparallel extension of the standard $\Lambda$CDM, which is recovered when $\beta = 0$. As can be seen by comparing Eq.~\eqref{constraintL} with Eq.~\eqref{eq:constraint}, the inclusion of a positive or negative cosmological constant shifts the curve in Fig.~\ref{fig:evo_beta_2} downwards or upwards, respectively, leading to lower or higher $\Omega_{\rm m0}$ values for a given $\beta$.

Following the same procedure as in Sec.~\ref{sec:6p}, and using Eqs.~\eqref{eq:FRL} and~\eqref{constraintL}, 
we now express the redshift as a function of the Hubble parameter as follows:
\begin{equation}
    z(H)=\left[\frac{(H^2/H_0^2-2\beta )e^{\beta H_0^2/H^2}-\Omega_{\Lambda0}}{(1-2\beta)e^{\beta}-\Omega_{\Lambda0}}\right]^{1/3}-1.
    \label{eq:z_L}
\end{equation}
The inclusion of $\Lambda$ in the action modifies the redshift dependence of the Hubble parameter, as seen in~Eq.~\eqref{eq:z_L}, and consequently changes the relation in Eq.~\eqref{eq:dH_dN} to the following form:
\begin{equation}
    \frac{dH}{d\mathcal{N}}=-\frac{3H}{2}\, \frac{1-(2\beta+\Omega_{\Lambda0}\,e^{-\beta H_0^2/H^2}) H_0^2/H^2}{1 -\beta  H_0^2/ H^2+2 \beta ^2 H_0^4/H^4}.\label{eq:dH_dN_L}
\end{equation}
 Notably, the denominator remains unchanged, preserving the property of not vanishing for any real values of $\beta$ and $H_0/H$, as discussed earlier in the context of Eq.~\eqref{eq:dH_dN}. The inclusion of $\Lambda$ affects the deceleration parameter, now given by
\begin{equation} \label{eq:q_L}
q=-1+\frac{3}{2}\, \frac{1-(2\beta+\Omega_{\Lambda0}\,e^{-\beta H_0^2/H^2}) H_0^2/H^2}{1 -\beta  H_0^2/ H^2+2 \beta ^2 H_0^4/H^4}.
\end{equation}
We define the combination of the terms arising from the $f(T)-T$ modification and the inclusion of a cosmological constant in the modified Friedmann equations as effective DE, with its energy density and pressure given by
\begin{align}
\label{eq:rhode}
    \rho_{\rm DE}&=\rho_{\rm T}+\rho_{\Lambda}, \\
    p_{\rm DE}&=p_{\rm T}+p_{\Lambda}, \label{eq:pde}
\end{align}
where $p_\Lambda=-\rho_\Lambda=- \Lambda/\kappa$, and $\rho_{\rm T}$ and $p_{\rm T}$ are given by
\begin{align}
    \label{eq:rhoT_LL} \rho_{\rm T}=&\frac{3 H^2}{\kappa}\left[1-\left(1-2\beta H_0^2/H^2\right)e^{\beta H_0^2/H^2}\right],\\
    p_{\rm T}=&-\frac{3 H_{0}^{2}}{\kappa}\Bigg[\frac{  \beta(1+2 \beta  H_{0}^{2}/H^{2})}{1-\beta H_0^2/H^2+2\beta^2H_0^4/H^4}\nonumber\\
&-\Omega_{\Lambda 0}\left(1-\frac{ {\mathrm e}^{-\beta  H_{0}^{2}/H^{2}} }{1-\beta H_0^2/H^2+2\beta^2H_0^4/H^4}\right)\Bigg].
\label{eq:pT_LL}
\end{align}
  Here, the case $\rho_{\rm T}(\beta=0)=0=p_{\rm T}(\beta=0)$ corresponds to the standard $\Lambda$CDM model, albeit based on TEGR with a cosmological constant, as expected. Notably, the cosmological constant does not affect $\rho_{\rm T}(H)$, which remains in the same form as in Eq.~\eqref{rho}, but it does influence $p_{\rm T}(H)$, though it reduces to Eq.~\eqref{pres} when $\Omega_{\Lambda0}=0$. The expression in Eq.~\eqref{eq:pT_LL} depends on $\Omega_{\Lambda 0}$ because we replaced the quantity $\dot{H}$ in the general definition of $p_{\rm T}$ from Eq.~\eqref{eq:pT_gen}. This implies that in the presence of several other components (radiation, neutrinos, etc.), $p_{\rm T}$ will change due to the variation in $\dot{H}$. We also give the EoS parameter for the effective DE, $w_{\rm DE} \equiv p_{\rm DE}/\rho_{\rm DE}$, in its explicit form as
\begin{equation}
\begin{aligned}   
\label{wdedef}
    w_{\rm DE}=&-\frac{\beta (1+2\beta H_0^2/H^2)+\Omega_{\Lambda0} e^{-\beta H_0^2/H^2}}{1-\beta H_0^2/H^2+2\beta^2 H_0^4/H^4}\\
   &\times\frac{H_0^2/H^2}{1-(1-2\beta H_0^2/H^2)e^{\beta H_0^2/H^2}+\Omega_{\Lambda0}H_0^2/H^2}.
\end{aligned}
\end{equation}

\begin{figure*}[t!]
\par
\begin{center}
\hspace*{-1.mm}\includegraphics[trim =0mm  0mm 0mm 0mm, clip, width=0.33\textwidth]{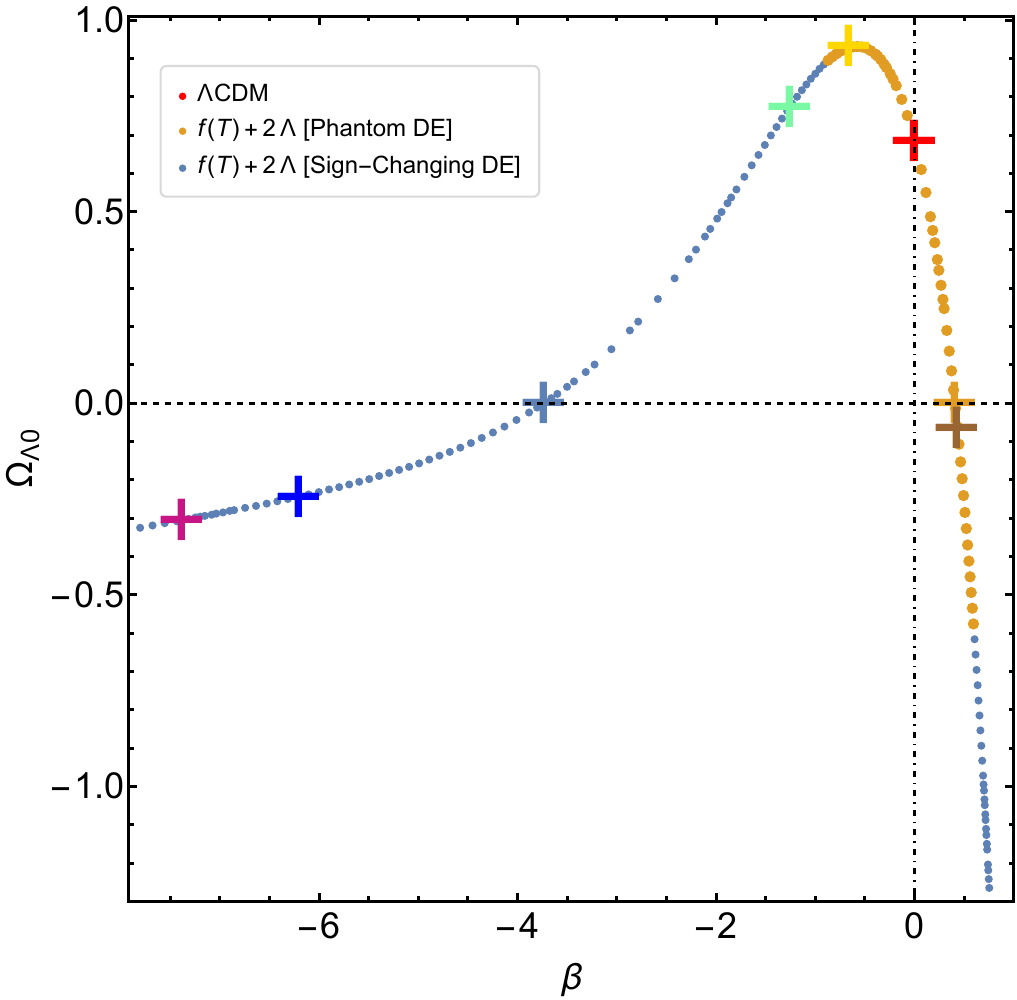} 
\hspace*{.5mm}\includegraphics[trim =0mm  0mm 0mm 0mm, clip, width=0.32\textwidth]{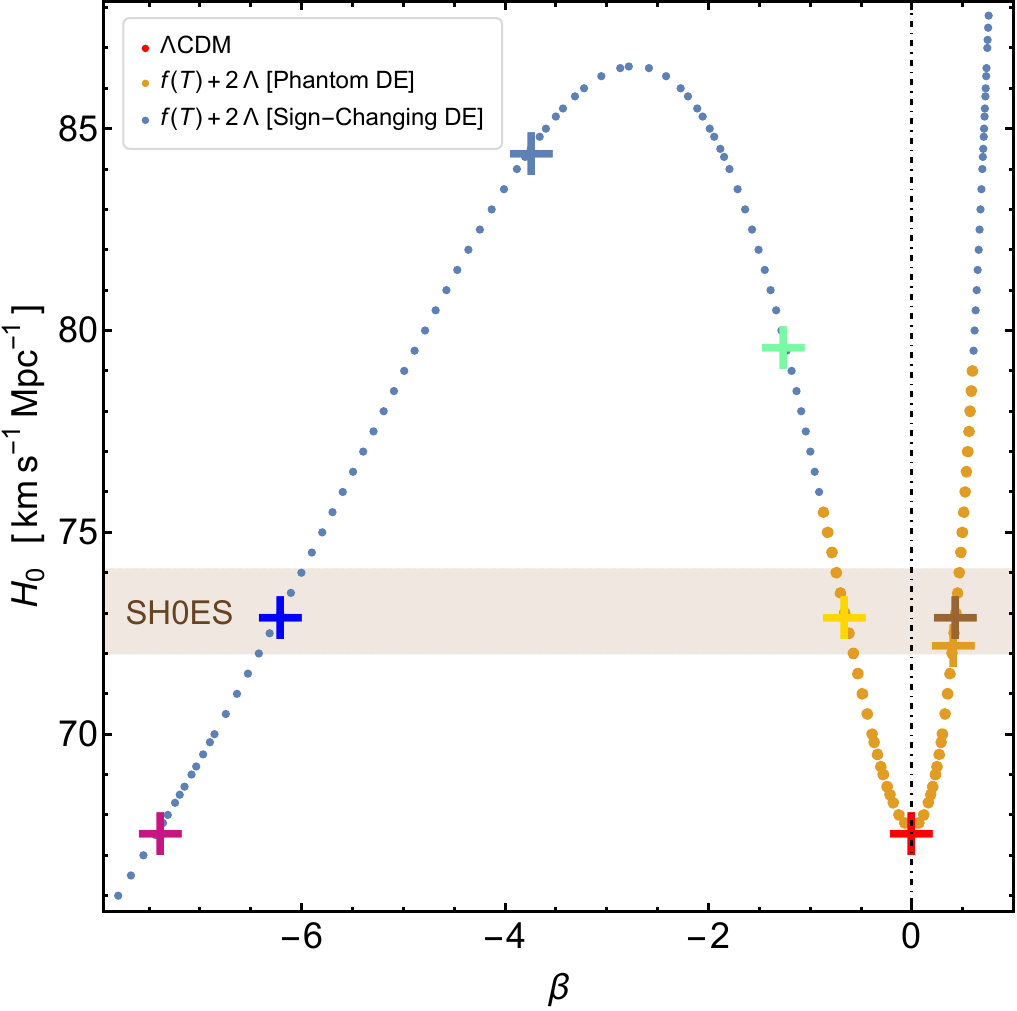} 
\includegraphics[trim = 0mm  0mm 0mm 0mm, clip, width=0.32\textwidth]{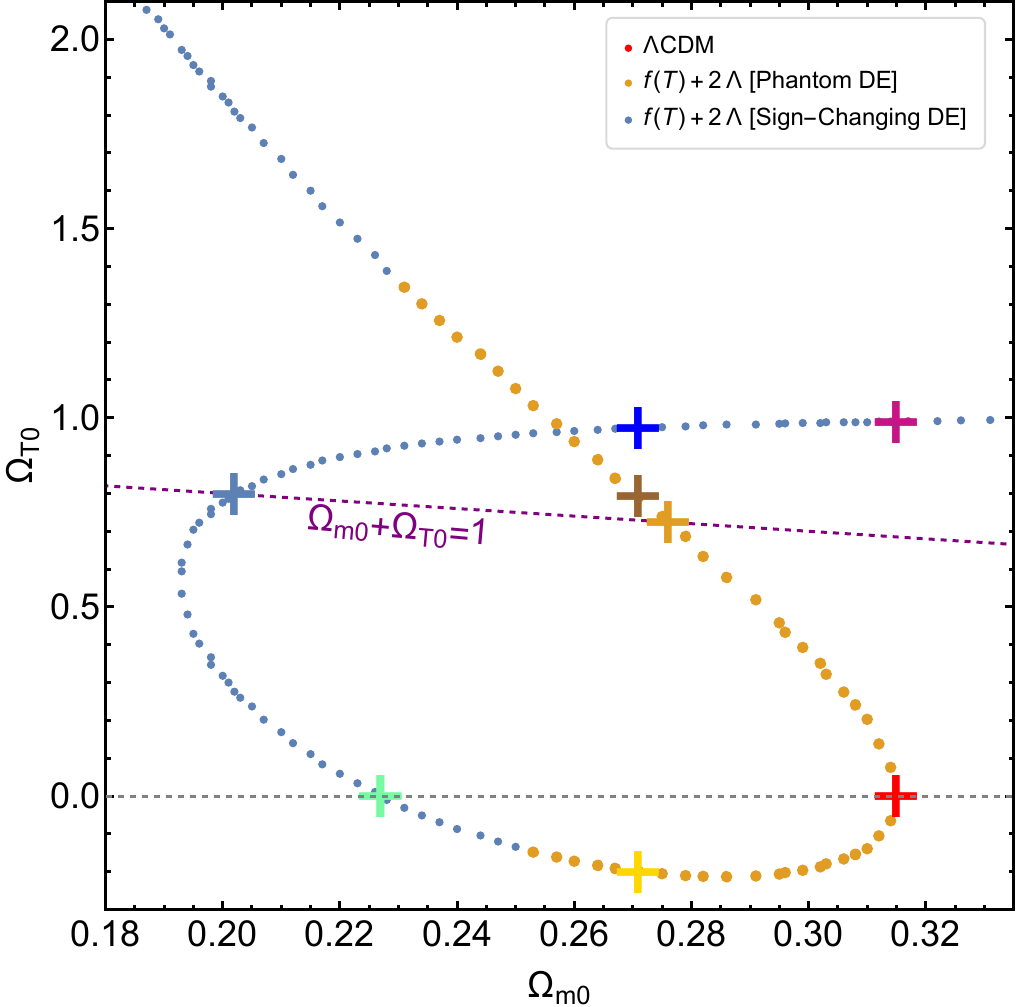}
\includegraphics[trim =0mm  0mm 0mm 0mm, clip, width=0.33\textwidth]{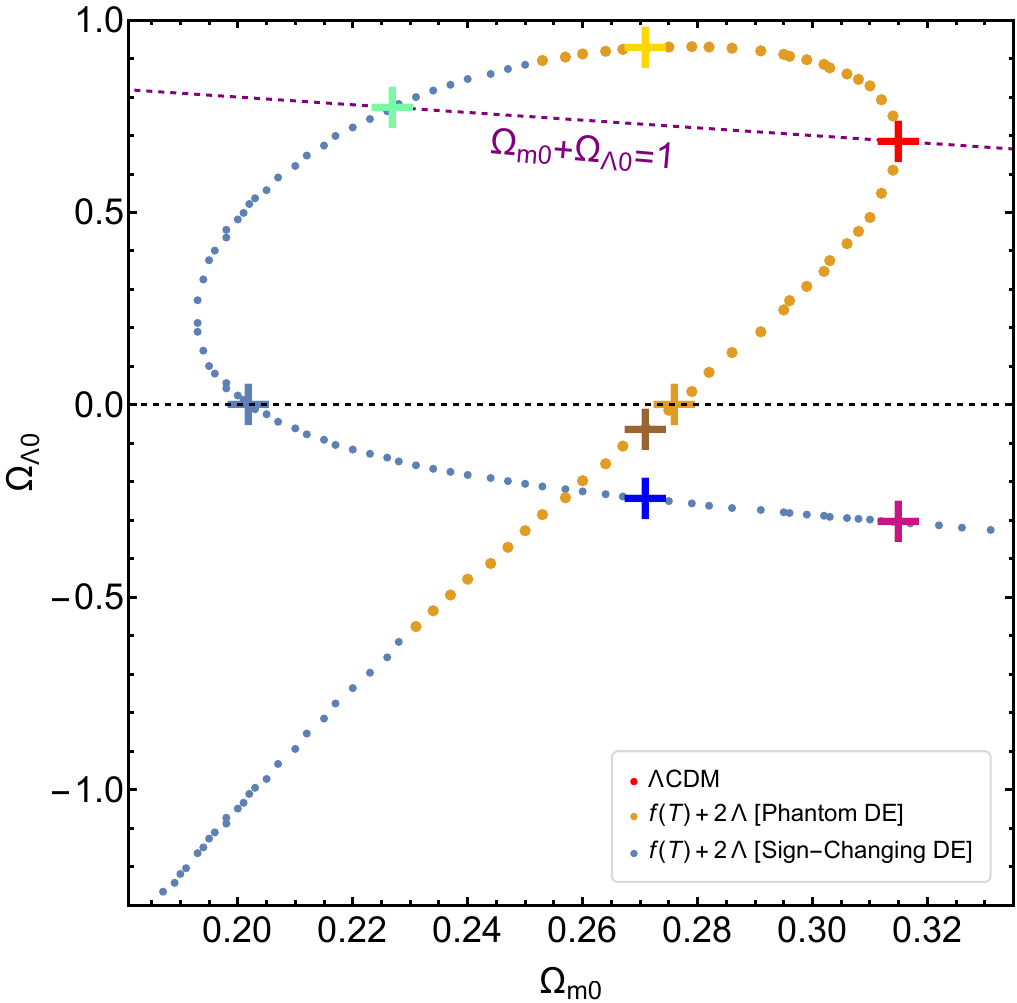}
\includegraphics[trim =0mm  0mm 0mm 0mm, clip, width=0.33\textwidth]{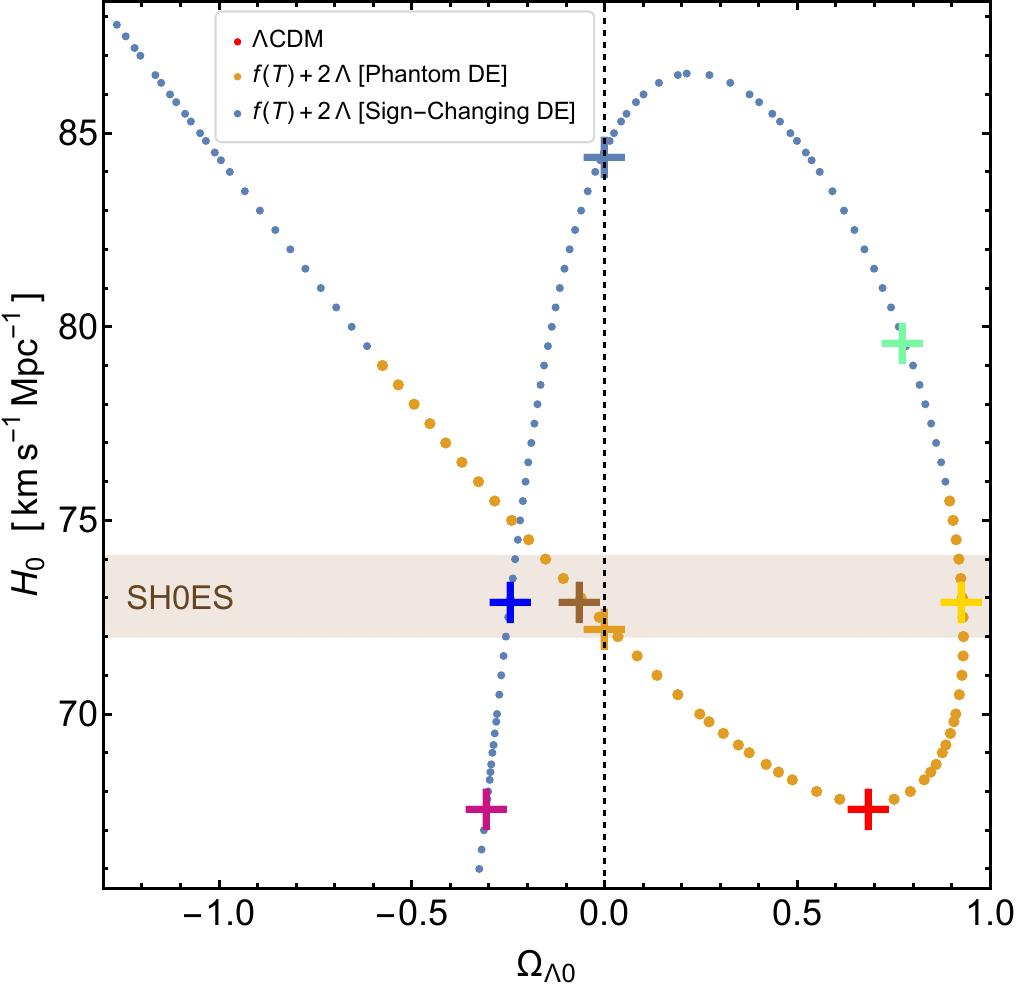}
\hspace*{-1.mm}\includegraphics[trim = 0mm  0mm 0mm 0mm, clip, width=0.328\textwidth]{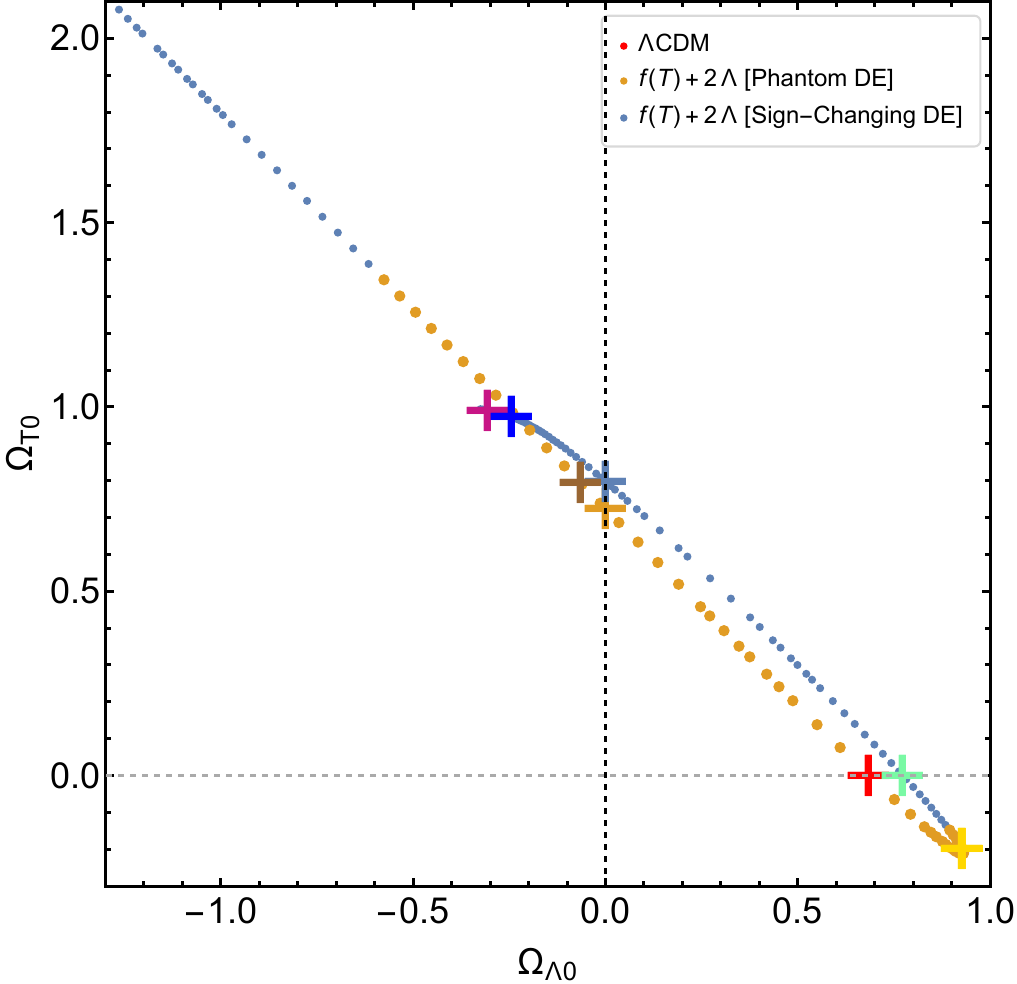}
\end{center}
\caption{\textbf{Top left panel:} $\Omega_{\Lambda0}$ vs $\beta$ graph. \textbf{Top middle panel:} $H_0$ vs $\beta$ graph.  \textbf{Top right panel:} $\Omega_{\rm T0}$ vs $\Omega_{\rm m0}$ graph. \textbf{Bottom left panel:} $\Omega_{\Lambda0}$ vs $\Omega_{\rm m0}$ graph. \textbf{Bottom middle panel:} $H_0$ vs $\Omega_{\Lambda0}$ graph. \textbf{Bottom right panel:} $\Omega_{\rm T0}$ vs $\Omega_{\Lambda0}$ graph, for exponential infrared teleparallel model with cosmological constant $\Lambda$. The dotted purple lines in the top right and bottom left panels represent $\Omega_{\rm m0}+\Omega_{\rm T0}=1$ and $\Omega_{\rm m0}+\Omega_{\Lambda0}=1$, respectively. The points correspond to models consistent with the CMB power spectra having parameter sets $(H_0, \Omega_{\rm m0}, \Omega_{\rm \Lambda0}, \beta)$ that satisfy the constraints on $D_{M}(z_*)$ and $\Omega_{\rm m0}h^2$ from \textit{Planck}-CMB observations. Orange points represent phantom DE models and blue points represent sign-changing DE models. Constraints at $1\sigma$ C.L. from SH0ES Collaboration measurement $H_0=73.04 \pm 1.04 \,{\rm km\,s}^{-1}{\rm Mpc}^{-1}$~\cite{Riess:2021jrx} are shown as a horizontal wheat-colored band. 
}
\label{Lambda-beta}
\end{figure*}

Having derived the essential equations and quantities for the extended model by including a cosmological constant $\Lambda$, we now proceed to analyze its richer phenomenological possibilities. The panels in Fig.~\ref{Lambda-beta} display the relationships between various parameter pairs: $\Omega_{\Lambda0} \, {\rm vs} \, \beta$ (top left), $H_0\, {\rm vs} \,\beta$ (top middle), $\Omega_{\rm{T0}}\, {\rm vs} \, \Omega_{\rm m0}$ (top right), $\Omega_{\Lambda0} \, {\rm vs} \, \Omega_{\rm m0}$ (bottom left), $H_0 \, {\rm vs} \,\Omega_{\Lambda0}$ (bottom middle), and $\Omega_{\rm{T0}} \, {\rm vs} \, \Omega_{\Lambda0}$ (bottom right). These results were obtained using Eqs.~\eqref{constraintL} and~\eqref{eq:z_L}, employing the same method presented in the previous section to ensure the consistency with the CMB data. Each point in the plots of Fig.~\ref{Lambda-beta} corresponds to a distinct cosmological model whose parameters satisfy the nearly model independent values of $D_{M}(z_*)= 13869.6 {\;\rm Mpc}$ and $\Omega_{\rm m0}h^2=0.14314$ (though replaced with $\Omega_{\rm m0}h^2=0.1444$ to account for the exclusion of radiation in the equations; see Sec.~\ref{sec:cosm} for a detailed explanation). Note that the relationships between other parameter pairs are straightforward; the $H_0 \, {\rm vs} \,\Omega_{\rm m0}$ relation is determined simply by the fixed physical energy density, meaning that replacing the $H_0$ axis with the $\Omega_{\rm m0}$ axis, or vice versa, effectively flips the plot---this can be observed by comparing the bottom left and middle panels. Additionally, the present-day density parameter of torsional DE is given by $\Omega_{\rm T0}=1-(1-2 \beta)e^{\beta}$ from Eq.~\eqref{constraintL}, ensuring that $\Omega_{\rm m0}+\Omega_{\Lambda0}+\Omega_{\rm T0}=1$. Our analysis is confined to the interval $(66 \leq H_0 \leq 88)\,{\rm km\,s}^{-1}{\rm Mpc}^{-1}$. Orange points correspond to an effective DE with a phantom EoS, similar to the model with $\beta_{+}$ discussed in Sec.~\ref{sec:phantom}, while blue points exhibit sign-changing behavior in the energy density of effective DE, reminiscent of the model with $\beta_{-}$ considered in Sec.~\ref{sec:signchange}. Remaining consistent with the color scheme from the previously discussed sections, we mark the phantom torsional DE model ($\beta_{+}, \Omega_{\Lambda0}=0$) with an orange plus sign, the sign-changing (in energy density) torsional DE model ($\beta_{-}, \Omega_{\Lambda0}=0$) with a blue plus sign, and the $\Lambda$CDM model with a red plus sign in the panels of Fig.~\ref{Lambda-beta}. The orange and blue plus signs are located on the $\Omega_{\Lambda0}=0$ lines, shown by dotted black lines in all panels, corresponding to the purple $\Omega_{\rm m0}+\Omega_{\rm T0}=1$ line in the top right panel, as expected. Similarly, the red plus sign ($\Lambda$CDM), where torsional effects vanish, is located on the $\beta=0$ (dot-dashed black) and $\Omega_{\rm T0}=0$ (dotted gray) lines, corresponding to the purple $\Omega_{\rm m0}+\Omega_{\Lambda0}=1$ line in the bottom left panel. Notably, $\Lambda$CDM is a special model, as the red plus sign corresponds to a local minimum in both the top and bottom middle panels.

\begin{figure*}[t!]
\par
\begin{center}
\includegraphics[trim =0mm  0mm 0mm 0mm, clip, width=0.48\textwidth]{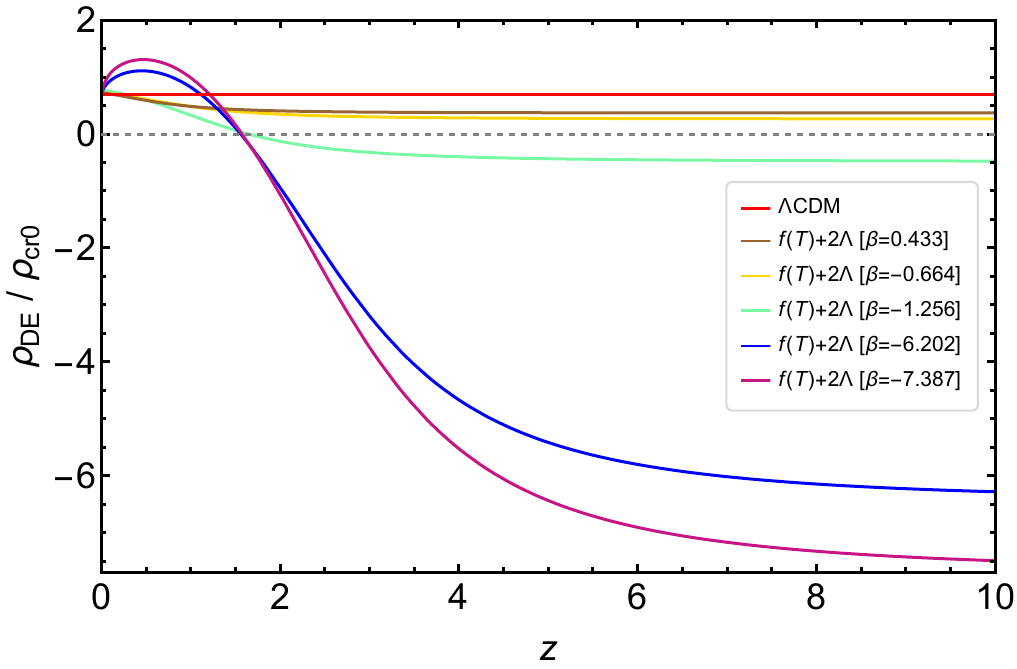}
\includegraphics[trim =0mm  0mm 0mm 0mm, clip, width=0.48\textwidth]{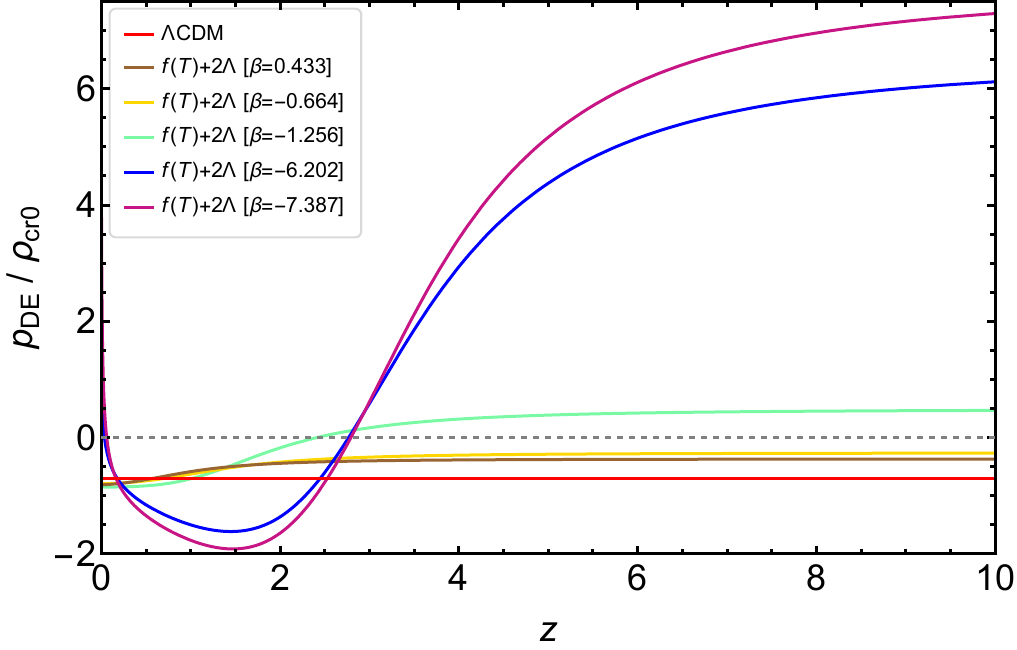} 
\hspace*{-3.mm}\includegraphics[trim =0mm  0mm 0mm 0mm, clip, width=0.49\textwidth]{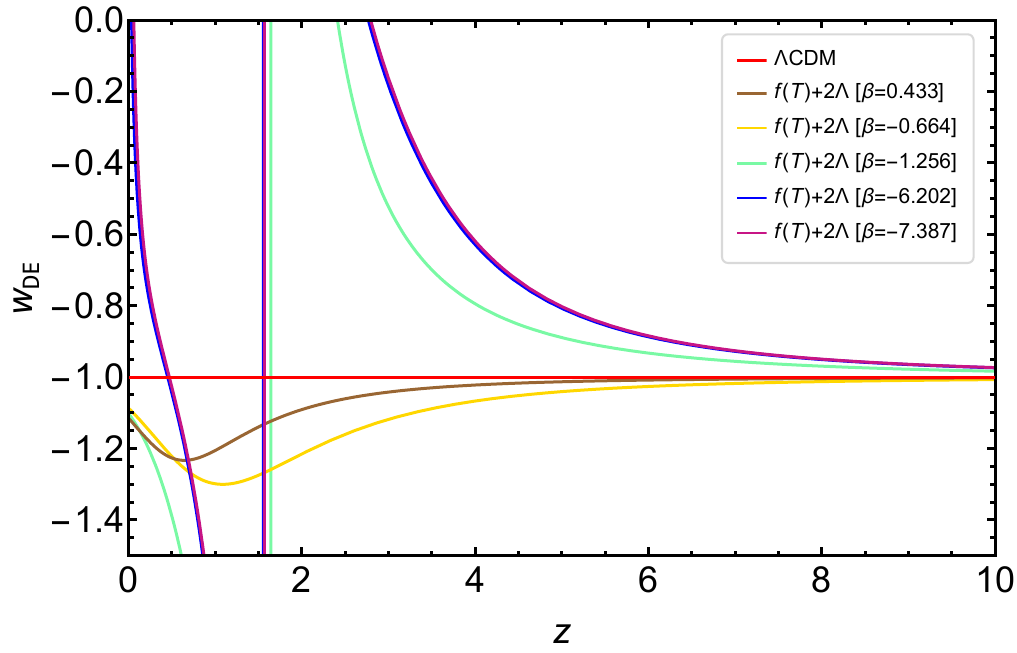}
\includegraphics[trim = 0mm  0mm 0mm 0mm, clip, width=0.48\textwidth]{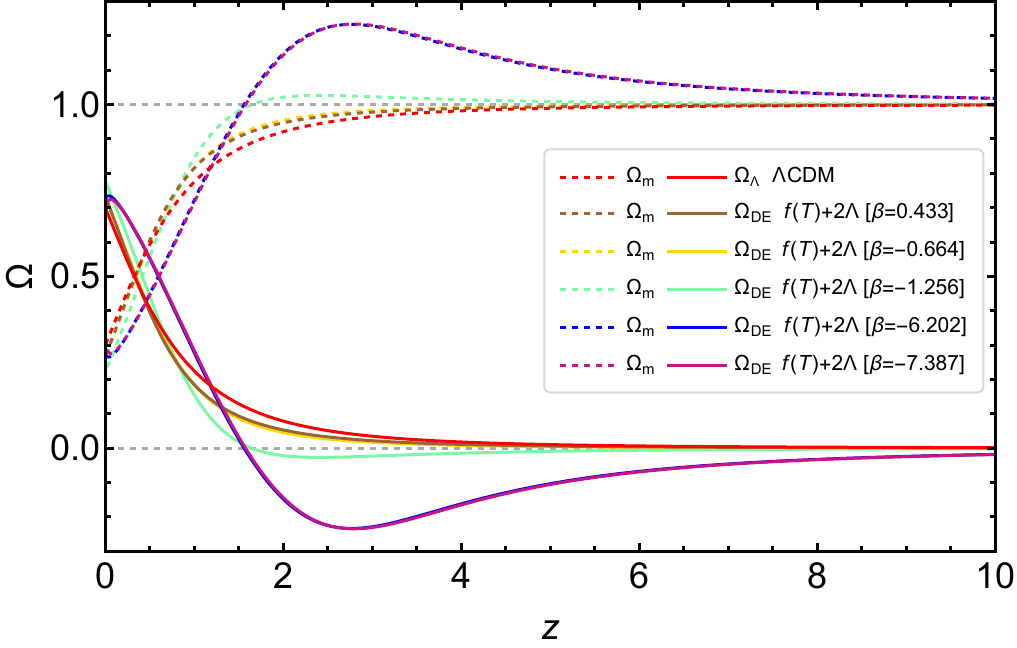}
\end{center}
\caption{\textbf{Top left panel:}  $\rho_{\rm DE}(z)/\rho_{\rm cr0}$ (energy density of DE scaled by critical density today). \textbf{Top right panel}: $p_{\rm DE}(z)/\rho_{\rm cr0}$ (pressure of DE scaled by critical density today). \textbf{Bottom left panel}:  $w_{\rm DE}(z)$ (EoS parameter of DE). \textbf{Bottom right panel}: $\Omega(z)$ (density parameters) of matter (dashed curves) and DE (solid curves). Colors are in agreement with those of plus signs used in Fig.~\ref{Lambda-beta}. }
\label{fig:wde,deltawde-L}
\end{figure*}

We now turn to the consequences of including $\Lambda$ in the exponential infrared model in terms of viable cosmologies, focusing on the characteristic parameters $\beta$ and $\Omega_{\Lambda0}$. It is important to emphasize that with the inclusion of $\Lambda$, the effective DE, defined in Eqs.~\eqref{eq:rhode} and~\eqref{eq:pde} as the combination of torsional DE (characterized by $\beta$) and the cosmological constant $\Lambda$, changes the classification based on the sign of $\beta$ used in the previous section. In other words, we observe that there are both orange and blue points on either side of the $\beta=0$ line in the top left and middle panels, indicating that it is now possible to realize phantom models with $\beta < 0$ and sign-changing energy density models with $\beta > 0$, in contrast to the $\Lambda = 0$ case.  It can be observed from the same panels that the orange points are distributed in a parabola-like shape around the $\beta=0$ line, and even for $-1 \lesssim \beta < 0$, with the help of a positive $\Lambda$ ($0.7 \lesssim \Omega_{\Lambda0} \lesssim 1$), the energy density of the effective DE remains positive without any sign change. On the other hand, in the same plots, all points are blue for $\beta \lesssim -1$, indicating that the negative value of $\beta$ dominates over the positive $\Lambda$ in the range $-4 \lesssim \beta \lesssim -1$, causing the energy density of effective DE to become negative beyond a certain redshift. Conversely, in the positive $\beta$ region, the dominance of a large negative $\Lambda$ for $\beta \gtrsim 0.6$ leads to sign-changing behavior. We observe that the points above the $\Omega_{\rm m0}+\Omega_{\Lambda0}=1$ line in the bottom left panel are below the $\Omega_{\rm T0}=0$ (dotted gray) line in the right panels, or vice versa, as expected since all three parameters must sum to unity. Similarly, the points above the $\Omega_{\rm m0}+\Omega_{\rm T0}=1$ line in the top right panel are below the $\Omega_{\Lambda0}=0$ (dotted black) line in the bottom left panel and to the left of the same line in the bottom right panel. The bottom left and top right panels show that for relatively small present-day matter density parameters ($0.18 \lesssim \Omega_{\rm m0} \lesssim 0.23$) and for relatively large ones ($0.315 \lesssim \Omega_{\rm m0} \lesssim 0.335$), no phantom models, i.e., orange points, are present. Comparing the bottom right and top left panels reveals a linear relationship between $\Omega_{\rm T0}$ and $\Omega_{\Lambda0}$ when $|\beta| \lesssim 0.7$. Next, we consider the horizontal wheat-colored band representing the SH0ES $H_0$ measurement, which corresponds to $(72.00 \leq H_0 \leq 74.08)\,{\rm km\,s}^{-1}{\rm Mpc}^{-1}$~\cite{Riess:2021jrx}. The top middle panel of Fig.~\ref{Lambda-beta} shows that when the Hubble constant is constrained within this range, the number of distinguishable phantom models is nearly double that of the sign-changing models. Phantom models are found around $\beta \approx -0.7$ and $\beta \approx 0.4$, whereas sign-changing models appear around $\beta \approx -6.5$, with no sign-changing models for $\beta > 0$ within the wheat-colored band. As shown by the orange points in the bottom middle panel of Fig.~\ref{Lambda-beta}, phantom models with negative $\Lambda$ have a broader horizontal intersection with the wheat-colored band compared to those with positive $\Lambda$. In other words, phantom DE with $\Lambda > 0$ is less likely than phantom DE with $\Lambda < 0$ when $H_0$ is constrained within the SH0ES measurement limits. For the phantom models, we find $\Omega_{\Lambda0} \approx 0.93$ for negative $\beta$ and $-0.2 \lesssim \Omega_{\Lambda0} \lesssim 0.04$ for positive $\beta$. However, for sign-changing models, the only viable value is $\Omega_{\Lambda0} \approx -0.24$ in the region intersecting with the wheat-colored band, as shown in the bottom middle panel. A Bayesian analysis would be necessary to compare these models with the standard $\Lambda$CDM model to determine which is more strongly supported by the datasets. This is especially relevant given that the model with negative $\beta$ is considered theoretically more elegant, as $\beta < 0$ ensures that $f_T$ remains positive, while $\beta > 0$ does not guarantee positivity throughout the Universe's evolution.

Among the viable model families consistent with the CMB power spectra, we highlight five special models in the panels of Fig.~\ref{Lambda-beta}, marked with brown, yellow, green, dark blue, and violet plus signs. In addition to $\Lambda$CDM (red plus sign), we observe in the bottom left panel that the $\Omega_{\rm m0}+\Omega_{\Lambda0}=1$ line is also intersected by the green plus sign, indicating another model where the present-day torsional contribution to the Friedmann equation vanishes (i.e., $\Omega_{\rm T0}=0$), as shown by the dotted gray line in the top and bottom right panels, with $\Omega_{\Lambda0}=0.773$ and $\beta=-1.256$ (top left panel). Featuring sign-changing behavior, its Hubble constant prediction, $H_0=79.74\,{\rm km\,s}^{-1}{\rm Mpc}^{-1}$, is significantly above the SH0ES measurement limits, as shown in the middle panels. The brown, yellow, and dark blue plus signs denote models whose Hubble constants perfectly match the SH0ES $H_0$ measurement~\cite{Riess:2021jrx}, with the first two representing phantom models and the last one representing a sign-changing model. At this point, we remind readers that the phantom model with $\beta_{+}$ and $\Omega_{\Lambda0}=0$ (orange plus sign) already predicts an $H_0$ value in strong agreement with the SH0ES measurement, and the inclusion of a cosmological constant appears to fine-tune the phantom models. However, the model marked by the yellow plus sign ($\beta=-0.664$, $\Omega_{\Lambda0}=0.928$) differs from those marked by the orange and brown plus signs ($\beta=0.433$, $\Omega_{\Lambda0}=-0.0649$) in that it achieves phantom behavior with a negative $\beta$. Also, note that the yellow plus sign is positioned very close to the local maximum in the top left panel and the local minimum in the bottom right panel. On the other hand, introducing $\Lambda$ can reduce the overestimated $H_0$ value of the sign-changing model with $\beta_{-}$ and $\Omega_{\Lambda0}=0$ (blue plus sign) to align with the SH0ES $H_0$ measurement, as demonstrated by the model marked by the dark blue plus ($\beta=-6.202$, $\Omega_{\Lambda0}=-0.244$), where the effect of the large negative $\beta$ is moderated by the negative value of $\Lambda$.  

Another notable feature of this sign-changing model is that it likely aligns well with some findings from the recently released DESI BAO data, as it suggests that the Universe's acceleration slows down at very low redshifts and even transitions into a decelerated expansion phase—a behavior distinct from the phantom models, which we will discuss further in this section. Similarly, there is another sign-changing model, marked by a violet plus sign ($\beta=-7.387$, $\Omega_{\Lambda0}=-0.305$), which matches the Hubble constant of \textit{Planck}-$\Lambda$CDM and exhibits similar behavior, with a shift to decelerated expansion at low redshifts.

\begin{figure}[t!]
\par
\begin{center}
\hspace*{-2.mm}
\includegraphics[trim =0mm  0mm 0mm 0mm, clip, width=0.48\textwidth]{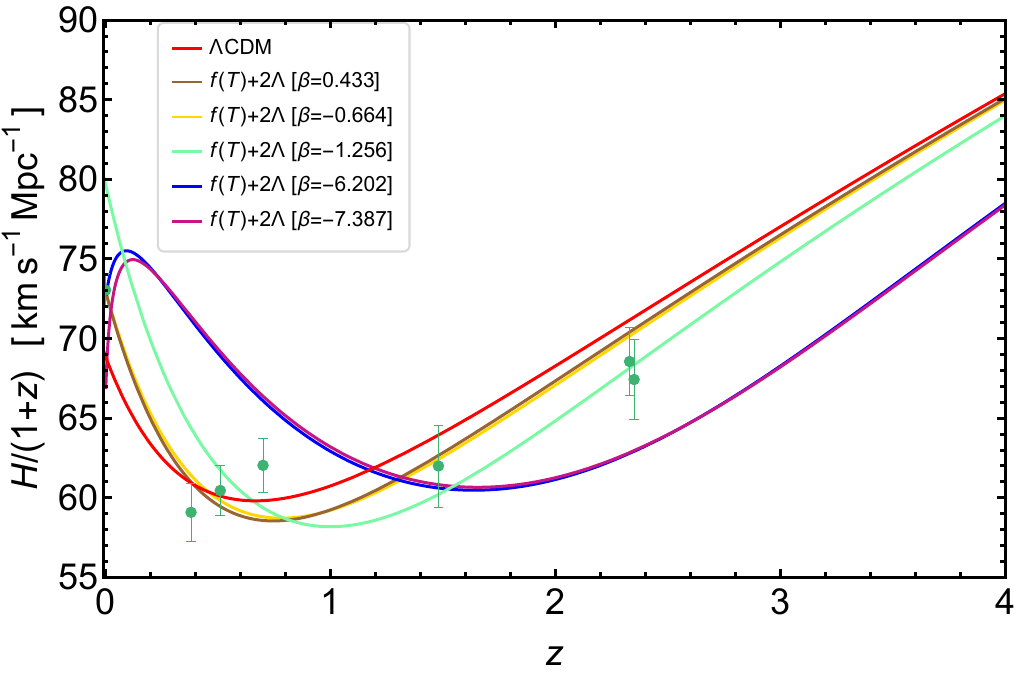} 
\hspace*{-1.mm}\includegraphics[trim = 0mm  0mm 0mm 0mm, clip, width=0.48\textwidth]{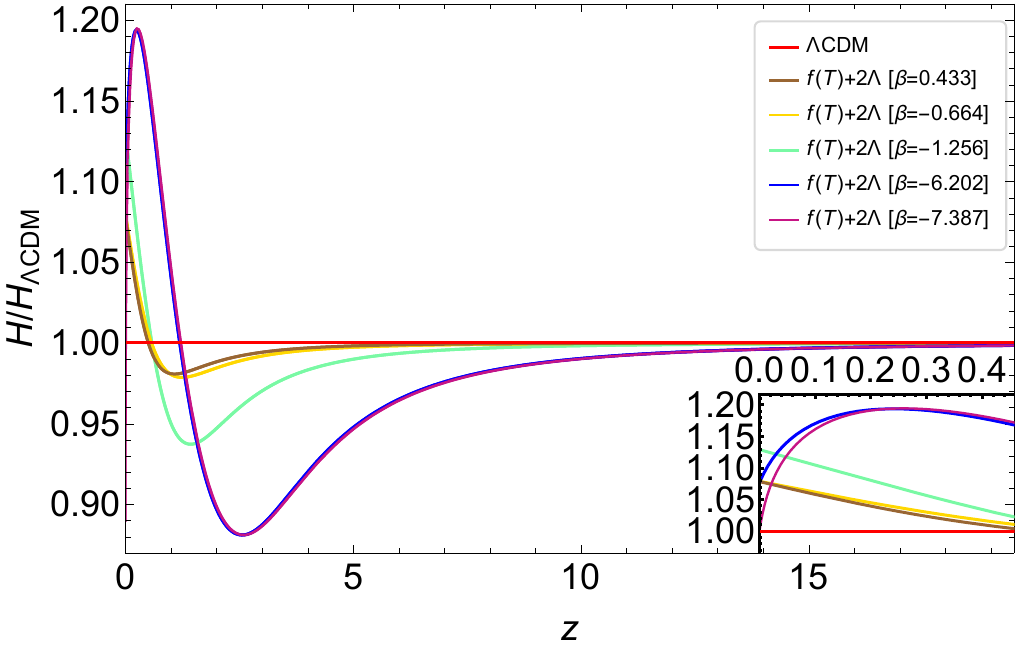}
\hspace*{-1.mm}\includegraphics[trim =0mm  0mm 0mm 0mm, clip, width=0.485\textwidth]{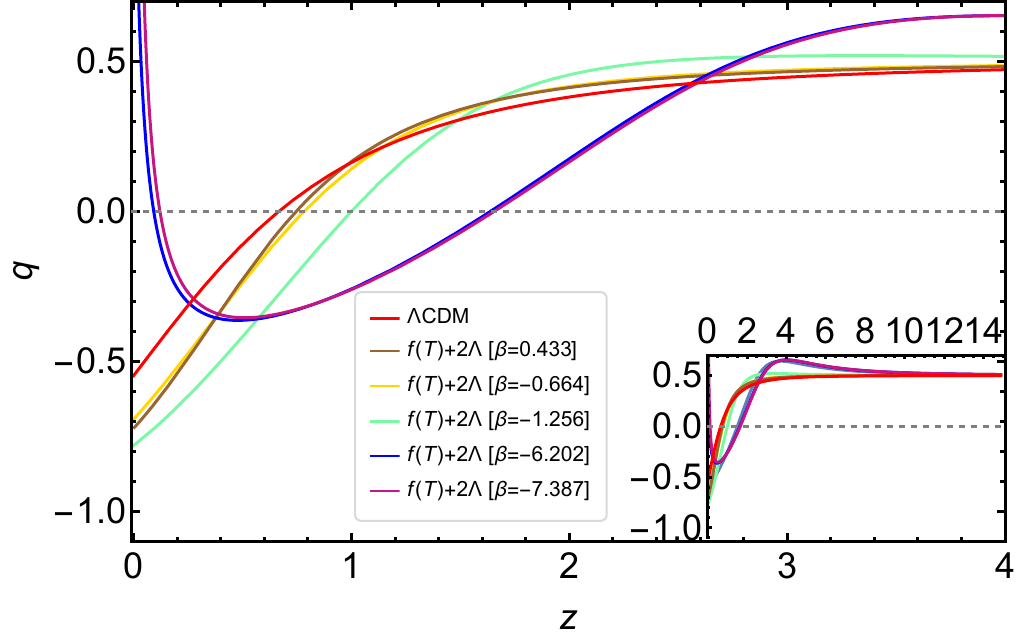}
\end{center}
\caption{\textbf{Top panel:} $\dot{a}=H(z)/(1+z)$ (comoving Hubble parameter, viz., expansion speed).  \textbf{Middle panel:} $H(z)/H_{\Lambda {\rm CDM}}(z)$. \textbf{Bottom panel:} $q(z)$ (deceleration parameter).
The green bars corresponds to SH0ES Collaboration measurement~\cite{Riess:2021jrx} and clustering measurements for the BAO samples in Ref.~\cite{eBOSS:2020yzd}; BOSS DR12 consensus Galaxy (from $z_{\rm{eff}} = 0.38, 0.51$), eBOSS DR16 LRG (from $z_{\rm{eff}}= 0.70$), eBOSS DR16 Quasar (from $z_{\rm{eff}}= 1.48$), eBOSS DR16 Ly$\alpha$-Ly$\alpha$ (from $z_{\rm{eff}} = 2.33$) and eBOSS DR16 Ly$\alpha$-quasar (from $z_{\rm{eff}} = 2.33$ but shifted to $z_{\rm{eff}} = 2.35$ in the figures for visual clarity) measurements.}
\label{fig:dec-comoving}
\end{figure}

Now, we turn to the detailed late-time dynamics of the five specific models marked by plus signs in Fig.~\ref{Lambda-beta}, as illustrated by a series of plots in Figs.~\ref{fig:wde,deltawde-L} and~\ref{fig:dec-comoving}. Comparing with the orange curves in Fig.~\ref{fig:wde,deltawde} from the previous section, we observe that the phantom models exhibit similar behavior in both the $\Omega_{\Lambda0} = 0$ and $\Omega_{\Lambda0} \neq 0$ cases, aligning with our earlier discussion on the fine-tuning effect of $\Lambda$ on these models. For the sign-changing models, we obtain $z_{\dagger}= 1.64$ for the green curve, $z_{\dagger}= 1.55$ for the dark blue curve, and $z_{\dagger}=1.57$ for the violet curve. These three redshifts of sign transition in DE density are quite close to that of $\Omega_{\Lambda0}=0$ case ($z_{\dagger}=1.45$). However, it is worth noting that there will be a second pole (singularity) in the future behavior of $w_{\rm DE}(z)$ (when $-1 < z < 0$) for the dark blue and violet curves, indicating that the DE density will cross zero, this time from positive to negative values, with the onset of this phase visible in the left panels of Fig.~\ref{fig:wde,deltawde-L}; note that the energy density tends to decrease while the EoS parameter increases asymptotically as $z \rightarrow 0$. More importantly, this nontrivial behavior paves the way for a slowing down in cosmic acceleration, reflected in the turning from increase to decrease in the comoving Hubble parameter, $H(z)/(1+z)$, in the neighborhood of $z=0$, as clearly illustrated in the panels of Fig~\ref{fig:dec-comoving}. This shows an astonishing parallelism with the findings in the model-agnostic DE reconstructions, using the Chebyshev expansion of the EoS parameter up to four terms, based on the latest BAO data from DESI, as reported in Calderon \textit{et al}.~\cite{DESI:2024aqx}, suggesting that the Universe has recently ceased accelerated expansion and today we are once again in a decelerated expansion phase; compare the behavior of the deceleration parameter $q(z)$ for $\beta=-7.878$ (dark blue curve) and $\beta=-6.202$ (violet curve) in the bottom panel of Fig.~\ref{fig:dec-comoving} with that shown in Fig.~1 of the DESI-BAO analysis. See also the earlier work by Ref.~\cite{Shafieloo:2009ti}, which suggests similar dynamics for the Universe at very low redshifts; namely, a coasting model of the Universe, with $q_0 \sim 0$, was shown to fit the data (SNIa+BAO data at the time) about as well as the $\Lambda$CDM model. Specifically, we observe an increase in $q(z)$ at redshifts $z \lesssim 0.5$, suggesting that cosmic acceleration may have already reached its peak and that we are currently witnessing a slowing down of this acceleration. This change in behavior leads to a prediction for the Hubble constant that aligns remarkably well with the SH0ES $H_0$ measurement of $H_0=73.04 \pm 1.04 \,{\rm km\,s}^{-1}{\rm Mpc}^{-1}$~\cite{Riess:2021jrx} for $\beta=-7.878$ (dark blue). On the other hand, for $\beta=-6.202$ (violet), the model predicts an $H_0$ consistent with the \textit{Planck}-$\Lambda$CDM value of $H_0=67.70 \,{\rm km\,s}^{-1}{\rm Mpc}^{-1}$. Despite the differences in $H_0$, in both cases, the behavior of $H(z)$ is significantly larger than that of the $\Lambda$CDM model over the redshift range $0.1 \lesssim z \lesssim 1$. This behavior emphasizes how the model's dynamics, especially its treatment of cosmic acceleration and deceleration, provide a framework that can match observational data, including both SH0ES and \textit{Planck} results, while exhibiting distinct differences from the standard $\Lambda$CDM model in the evolution of $H(z)$ at intermediate redshifts. Therefore, before concluding this section, let us take a closer look at these models. As observed from the bottom panel of Fig.~\ref{fig:dec-comoving}, the deceleration parameter reaches its minimum value and begins to increase around $z \sim 0.5$, eventually crossing zero ($q=0$), marking the turning point where the comoving Hubble parameter reaches its local maximum (expansion speed, i.e., $\dot{a}=H(z)/(1+z)={\rm const.}$), as seen in the top panel of the same figure. Specifically, this second sign transition of $q$—from acceleration to deceleration—in the post-recombination history of the Universe occurs in today's neighborhood at $z_{\rm tr}=0.09$ for the dark blue curve and $z_{\rm tr}=0.12$ for the violet curve, while the first transition occurred around $z\sim1.6$ for both models, marking the beginning of the Universe’s late-time accelerated expansion phase.  This leads to an oscillating-like behavior in $H(z)/H_{\Lambda {\rm CDM}}(z)$ at low redshifts. As shown in the middle panel of Fig.~\ref{fig:dec-comoving}, the ratio of the Hubble parameter in these models, $H(z)$, to that of the $\Lambda$CDM model, $H_{\Lambda \rm CDM}(z)$, oscillates between approximately $0.88$ and $1.20$ for both the dark blue and violet curves. Such oscillatory behaviors have been suggested with varying significance depending on the Chebyshev expansion of DE density or EoS, using DESI BAO data and its combinations with other datasets such as SNIa compilations from Union3, PantheonPlus, and Dark Energy Survey Supernova 
Program 5-Year results (DES-SN5YR), as reported in Ref.~\cite{DESI:2024aqx}. Similar kinematic deviations—oscillatory patterns at low redshifts and significantly smaller values of $H(z)$ for $z \gtrsim 1.5-2$ compared to the \textit{Planck}-$\Lambda$CDM model—have also been reported in other studies, particularly in model-independent or non-parametric reconstructions of the late Universe dynamics using different datasets and methods (see e.g., Refs.~\cite{Escamilla:2024ahl,Sabogal:2024qxs,Pace:2012oscillating,Tamayo:2019gqj,Akarsu:2022lhx, Rezaei:2024vtg,Escamilla:2024fzq,Escamilla:2023shf,Escamilla:2021uoj} and references therein). These deviations are often attributed, under the assumption of GR, to nontrivial DE dynamics, such as oscillatory EoS or energy density at low redshifts, and energy densities taking negative values at larger redshifts (e.g., for $z \gtrsim 1.5-2$). The fact that we obtain such dynamics for the late Universe within the $f(T)$ gravity framework, consistent with these independent model-agnostic observational studies, further motivates exploring unexplored regions of the $f(T)$ models in the literature or investigating new functions for $f(T)$ models that could realize nontrivial deviations from the standard $\Lambda$CDM model, while potentially fitting the data better and addressing cosmological tensions.

\section{Conclusion}
\label{sec:final}

In this paper, we have examined previously unexplored solution spaces within teleparallel $f(T)$ gravity, demonstrating that this promising theory, already capable of alleviating cosmological tensions, also has the potential to realize a sign change in the DE density—a transition from negative values in the past to positive values in the late Universe—to achieve even greater success. We have adopted the exponential infrared teleparallel model of gravity described by the gravitational Lagrangian density $f(T)=Te^{T_*/T}$—introduced in Ref.~\cite{Awad:2017yod}—and performed a theoretical analysis to explore all possible solution regions under the assumptions of the spatially flat FLRW metric, describing spacetime and a perfect fluid (dominantly pressureless matter at late times) filling the Universe. Defining the characteristic torsion scale as $T_*=\beta T_0$, where $T_0$ denotes the present-day value of the torsion scalar $T$ in this background, the $f(T)$ model under consideration is special in that the dimensionless parameter $\beta$ is not a free parameter but is instead determined by the present-day matter density parameter, $\Omega_{\rm m0}$, via a constraint equation, and therefore introduces no additional free parameters compared to the standard $\Lambda$CDM model. We have expressed the redshift dependencies of all quantities in parametric form, where the Hubble parameter $H$ acts as a free variable since it cannot be isolated in the modified Friedmann equation.

Depending on the value of the $\beta$ parameter, which represents a critical part of the exponent, we identify four distinct regions and three special points that exhibit different background dynamics, leading to various possibilities for the evolution of the Hubble and deceleration parameters within a theoretical framework. Among them, only Region II, involving positive exponents ($\beta_{+}$), and Region IV, involving negative exponents ($\beta_{-}$), are relevant when we narrow the interval for the present-day matter density parameter down to reasonable values, i.e., $0.2 \leq \Omega_{\rm m0} \leq 0.4$. We then treat the modified Friedmann equations as those of a dynamical DE model and interpret the extra torsional terms arising from the $f(T)$ modification as the energy density, $\rho_{\rm T}$, and pressure, $p_{\rm T}$, of effective torsional DE. In these regions, we obtain $\rho_{\rm T} \to \beta \rho_{\rm cr0}$ as $z \to \infty$, where $\rho_{\rm cr0}$ is the present-day critical density, meaning that the sign of $\beta$ determines the sign of the early-time energy density of torsional DE. Since the energy density of torsional DE, whose present-day value is positive, decreases with increasing redshift, this implies that for $\beta_{+}$, it remains positive throughout the history of the Universe, whereas for $\beta_{-}$, it evolves into negative values after passing through zero. On the other hand, different signatures of $\beta$ do not theoretically offer equal advantages; particularly, negative values of $\beta$ ensure that the derivative of $f(T)$ with respect to $T$ remains positive, viz., $f_{T} > 0$ for $\beta_{-}$, throughout, thereby guaranteeing the attractive nature of gravity and the absence of ghost instabilities. Previous studies~\cite{Awad:2017yod,Hashim:2020sez,Hashim:2021pkq}, however, focused only on models with $\beta_{+}$, resulting in a DE component whose effective equation of state (EoS) parameter is below $-1$, hence in the phantom phase, and excluded/overlooked models with $\beta_{-}$, possibly due to the conventional assumption that DE density remains positive (though this assumption may be safely relaxed for an effective DE from modified gravity). In contrast, Refs.~\cite{Akarsu:2019hmw,Akarsu:2021fol,Adil:2023exv} argue that a DE density that monotonically decreases, like usual phantom fields, but assumes negative values in the past may be considered a natural extension of phantom models and can achieve higher values of $H_0$ than phantom models confined to yield only positive energy densities. In line with this, we have shown that such nontrivial torsional DE, with a sign-changing feature in its density accompanied by a singularity in its EoS parameter, can be achieved in the exponential infrared teleparallel model for the case of $\beta_{-}$, even without requiring a cosmological constant. Furthermore, $\beta_{-}$ has a finite Minkowski limit in the future, viz., $H \to 0$ for $z \to -1$, distinguishing it from $\beta_{+}$, which approaches the dS limit in the future, as is familiar from the $\Lambda$CDM model, in addition to avoiding instabilities/ghosts. 

To ensure good consistency with the CMB power spectra, we have imposed constraints from \textit{Planck}-CMB observations on the comoving angular diameter distance $D_{M}(z_*)$ and the physical matter density $\Omega_{\rm m0}h^2$, and thus obtained the following parameters: $\Omega_{\rm m0}=0.276$ and $\beta_{+} = 0.408$; $\Omega_{\rm m0}=0.202$ and $\beta_{-}=-3.736$; and $\Omega_{\rm m0}=0.315$ for $\Lambda$CDM. The model with $\beta_{+}$ exhibits a phantom character at $0 \leq z \lesssim 5$, whereas it becomes almost indistinguishable from $\Lambda$ at larger redshifts, viz., $z \gtrsim 5$. This shifts the beginning of the late-time accelerated expansion phase of the Universe to earlier times compared to the one in the base $\Lambda$CDM model, and, similar to a domino effect, the pronounced phantom character of the torsional DE at low redshifts, in turn, gives rise to a higher value of the Hubble constant, $H_0=72.36\, {\rm km\,s}^{-1}{\rm Mpc}^{-1}$ (matching the constraints from observational analysis~\cite{Hashim:2021pkq}), in great agreement with the SH0ES $H_0$ measurement~\cite{Riess:2021jrx}, as well as providing a better description of the BAO data at high redshifts (viz., Ly$\alpha$ data) compared to the $\Lambda$CDM model. On the other hand, the torsional DE in the model with $\beta_{-}$ features an intervening phantom regime at $0.2 \lesssim z < 1.45$ (after the sign change) between two quintessence regions, residing in an unexplored solution region beyond the phenomenological possibilities studied so far. This case leads to an overenhanced Hubble constant, $H_0=84.54\, {\rm km\,s}^{-1}{\rm Mpc}^{-1}$, resulting from significantly boosted $H(z)$ values for $z \lesssim 1$ compensating for the notably low $H(z)$ values for $z \gtrsim 1$—caused by the large negative DE density—in the range $1.45 \lesssim z \lesssim 8$, in order to match the fixed value of the $D_{M}(z_*)$. Revealing the $\beta_{-}$ case provides insight into how this particular $f(T)$ model offers an effective framework for alleviating major cosmological tensions, although the resulting enhancement is greater than initially anticipated due to the strength of $\beta$, which is governed solely by $\Omega_{\rm m0}$.

Next, our findings led us to introduce an additional degree of freedom; we chose to include the well-known cosmological constant $\Lambda$, as the most straightforward option, which gives rise to an extended model corresponding to the replacement $f(T) \rightarrow f(T) + 2\Lambda$, where the torsion and $\Lambda$ work in tandem. Retaining $f(T) = T e^{\beta T_0/T}$ in the new model, which can be considered as a one-parameter exponential infrared teleparallel extension of the standard $\Lambda$CDM from an alternative perspective, the cosmological possibilities consistent with the CMB power spectra evolve into an incredibly broad and promising solution space (refer to Fig.~\ref{Lambda-beta}), part of which is in excellent agreement with the SH0ES $H_0$ measurement (points within the wheat-colored band in Fig.~\ref{Lambda-beta}) and whose $\beta = 0$ point now exactly corresponds to $\Lambda$CDM. The fact that $\Lambda$CDM resides in the local minimum in the $H_0 \, \rm{vs}\,\beta$ and $H_0 \, \rm{vs}\, \Omega_{\rm m0}$ plots suggests that extending with $\Lambda$ was a wise choice for alleviating the $H_0$ tension. We have noticed that, after introducing $\Lambda$, the classification based on the sign of $\beta$ used in Sec.~\ref{sec:cosm} is no longer applicable as it stands. The DE is now composed of both the cosmological constant and torsion, allowing for four distinct combinations based on the signs of $\beta$ and $\Omega_{\Lambda0}$. Among them, ($\beta < 0$ with $\Lambda < 0$) does not permit a phantom DE, and ($\beta > 0$ with $\Lambda > 0$) does not allow for a sign-changing DE, whereas the other two combinations accommodate both. It is clear that a Bayesian analysis is essential to compare these models, which exhibit distinct background dynamics, against the standard $\Lambda$CDM model to determine which is more strongly supported by the available data, particularly the $H_0$ value measured by the SH0ES Collaboration.

We have already seen that the phantom model with $\beta_{+}$ and $\Omega_{\Lambda0}=0$ predicts an $H_0$ value in strong agreement with the SH0ES measurement, and the inclusion of a cosmological constant indeed fine-tunes it in one of the models whose Hubble constant perfectly matches the SH0ES measured value. Additionally, another such model achieves phantom behavior with a negative $\beta$, and in this sense, it differs from these two. This model is positioned very close to the local maximum, having the largest $\Omega_{\Lambda0}$ accompanied by the smallest $\Omega_{\rm T0}$. In an interesting sign-changing model, although $\beta$ is nonvanishing, the torsional contribution vanishes today ($\Omega_{\rm T0}=0$) as in $\Lambda$CDM but diminishes to negative values with increasing redshift, while the large positive $\Omega_{\Lambda0}$ pulls the Hubble constant to higher values. We observed that the sign-changing model, where the effect of the large negative $\beta$ is moderated by the negative value of $\Lambda$, can reduce the overestimated $H_0$ value of the model obtained with $\beta_{-}$ and $\Omega_{\Lambda0}=0$ to match the SH0ES-measured value. In this model, the cosmic expansion undergoes a slowing down of acceleration at very low redshifts and even transitions into a decelerating phase---a distinguishing behavior not observed in phantom models, and akin to findings from the recent DESI BAO data analysis by Calderon \textit{et al}.~\cite{DESI:2024aqx}. This slowing down of acceleration, transitioning into deceleration at very low redshifts, triggers a decrease in the $H_0$, which would otherwise tend to prefer an overenhanced value. A similar behavior---a shift to a deceleration phase at low redshifts---also exists in another sign-changing model, though in this case, the Hubble constant matches that of \textit{Planck}-$\Lambda$CDM. In this model, as the redshift increases, the present-day positive torsional contribution first rises briefly before decreasing and becoming highly negative at earlier times. The sharp reduction in $H_0$ reflects the deceleration at very low redshifts, along with $\Omega_{\Lambda0} < 0$.

As mentioned earlier, the rapid cessation of cosmic acceleration at very low redshifts, viz., in the neighborhood of $z=0$, which manifests itself as a local maximum in the expansion speed, $\dot{a}$, observed in some sign-changing models within $f(T)$ framework, is quite similar to the behavior reported in the recent findings from the DESI BAO analysis by Calderon \textit{et al}~\cite{DESI:2024aqx}. This behavior is reflected in an oscillating-like pattern in the ratio of the Hubble parameter $H(z)$ to that of $\Lambda$CDM, $H_{\Lambda \rm{CDM}}(z)$, and a second sign transition---from acceleration to deceleration, in contrast to the first transition marking the beginning of the late-time accelerated expansion phase of the Universe---in the deceleration parameter $q(z)$. These previously overlooked, promising nontrivial background dynamics, along with the positivity of $f_T$ (ensuring that the effective cosmological Newtonian constant $\mathcal{G}_{\text{eff}}$ remains positive as well as the instabilities or ghosts are avoided) achieved by $\beta < 0$, encourage further research in $f(T)$ gravity, including revisiting the existing literature and exploring new functional forms that can give rise to these dynamics.  Another property inherent to these  models is that the effective Newtonian constant $\mathcal{G}_{\text{eff}}$, which governs cosmological perturbations, deviates significantly from the standard Newtonian constant $G_{\rm N}$ due to the large negative values of $\beta$ required to realize these scenarios, which imply a large $f_T$ connecting these constants. However, caution must be exercised in drawing conclusions, as the effective cosmological Newtonian constant $\mathcal{G}_{\text{eff}}$, does not necessarily coincide with the effective Newtonian constant $G_{\text{eff}}$ measured locally (e.g., within the Solar System), where we expect $G_{\text{eff}} \approx G_{\rm N}$, given both constants' dependence on the torsion scalar $T$, which is influenced by local spacetime properties. In other words, environmental factors influence the behavior of $f(T)$ gravity across different scales, potentially leading to distinct gravitational dynamics in low- and high-density regimes. In this context, we have also discussed that, despite the absence of an explicit scalar-matter coupling in pure $f(T)$ gravity, an \emph{effective chameleon-like mechanism} can still emerge under appropriate conditions, naturally suppressing deviations from GR in high-density environments and thereby ensuring consistency with local gravity tests. This reinforces the viability of negative-$\beta$ models, which—while exhibiting significant departures from $\Lambda$CDM at cosmological scales—can remain consistent with Solar-System tests and offer a promising path for realizing a sign-changing effective DE component. In particular, the previously unexplored region of $f(T)$ models with negative $\beta$ values on the order of $\mathcal{O}(1)$ might hold a significant advantage, enabling a cosmologically promising effective DE density that assumes negative values in the past and changes sign at low redshifts.

Thus, our findings suggest that $f(T)$ gravity holds substantial promise for explaining cosmological phenomena and effectively addressing major cosmological tensions, particularly when the customary assumption of strict positivity for the effective DE density is relaxed. We propose that future theoretical and observational studies revisit existing $f(T)$ cosmological models from this perspective,\footnote{As previously noted, in spatially flat FLRW spacetimes, the field equations of $f(T)$ gravity are exactly equivalent at the background level to those of $f(Q)$ gravity formulated in the coincident gauge with the appropriate connection~\cite{Jarv:2018bgs}. Accordingly, the results obtained in this work for the specific exponential $f(T)$ model~\cite{Awad:2017yod,Hashim:2020sez,Hashim:2021pkq} apply directly to its symmetric teleparallel $f(Q)$ counterpart, which has been explored in Refs.~\cite{Anagnostopoulos:2021ydo,Khyllep:2022spx,Lymperis:2022oyo,Boehmer:2023knj,Ferreira:2023awf,Yang:2024tkw,Boiza:2025xpn}. This further highlights the broader significance of our findings within the so-called “geometrical trinity of gravity” framework~\cite{BeltranJimenez:2019esp,Capozziello:2022zzh}, which encompasses curvature-, torsion-, and nonmetricity-based formulations of GR. It is therefore imperative to revisit such models---both with and without a cosmological constant $\Lambda$---to explore their largely uncharted parameter space and uncover their distinctive signatures at the perturbative level, while ensuring compatibility with CMB observations.}
 with the present work providing a foundational example of such an approach. Furthermore, exploring novel functional forms for $f(T)$ models (e.g., embedding $\Lambda_{\rm s}$CDM scenario into the $f(T)$ gravity~\cite{Souza:2024qwd}) could reveal unforeseen, nontrivial deviations from the standard $\Lambda$CDM model, offering improved data compatibility and success in addressing the persistent cosmological tensions.

\begin{acknowledgments}
\"{O}.A. acknowledges the support by the Turkish Academy of Sciences in the scheme of the Outstanding Young Scientist Award  (T\"{U}BA-GEB\.{I}P). The work of A.D.F. was supported by the Japan Society for the Promotion of Science Grants-in-Aid for Scientific Research No.~20K03969. This study was supported by Scientific and Technological Research Council of T\" urkiye (T\"{U}B\.{I}TAK) under Grant No.~122F124. The authors thank T\"{U}B\.{I}TAK for their support. N.M.U. was supported first by  T\"{U}B\.{I}TAK Grant No. 122F124, and then by T\"{U}B\.{I}TAK Grant No. 124C450 during this project. This paper is based upon work from COST Action CA21136 Addressing Observational Tensions in Cosmology with Systematics and Fundamental Physics (CosmoVerse) supported by COST (European Cooperation in Science and Technology).
\end{acknowledgments}  

\newpage

\bibliographystyle{apsrev4-2_mod}
\bibliography{references}
\end{document}